\documentclass[twocolumn,english,aps,prl,showpacs,superscriptaddress,preprintnumbers]{revtex4}
\usepackage[T1]{fontenc}
\usepackage[latin9]{inputenc}
\usepackage{float}
\usepackage{amsmath}
\usepackage{amssymb}
\usepackage{graphicx}
\usepackage{esint}

\makeatletter

\@ifundefined{textcolor}{}
{%
 \definecolor{BLACK}{gray}{0}
 \definecolor{WHITE}{gray}{1}
 \definecolor{RED}{rgb}{1,0,0}
 \definecolor{GREEN}{rgb}{0,1,0}
 \definecolor{BLUE}{rgb}{0,0,1}
 \definecolor{CYAN}{cmyk}{1,0,0,0}
 \definecolor{MAGENTA}{cmyk}{0,1,0,0}
 \definecolor{YELLOW}{cmyk}{0,0,1,0}
}

\usepackage{amstext}
\usepackage{babel}
\newcommand{\unit}{1\!\!1}
\addto\captionsenglish{}
\makeatother

\begin{document}

\title{Strong coupling of optical nanoantennas and atomic systems}
\author{K. S\l{}owik}
\affiliation{Institute of Condensed Matter Theory and Solid State Optics, Abbe
Center of Photonics, Friedrich-Schiller-Universit\"{a}t Jena, D-07743
Jena, Germany}
\author{R. Filter}
\affiliation{Institute of Condensed Matter Theory and Solid State Optics, Abbe
Center of Photonics, Friedrich-Schiller-Universit\"{a}t Jena, D-07743
Jena, Germany}
\author{J. Straubel}
\affiliation{Institute of Condensed Matter Theory and Solid State Optics, Abbe
Center of Photonics, Friedrich-Schiller-Universit\"{a}t Jena, D-07743
Jena, Germany}
\author{F. Lederer}
\affiliation{Institute of Condensed Matter Theory and Solid State Optics, Abbe
Center of Photonics, Friedrich-Schiller-Universit\"{a}t Jena, D-07743
Jena, Germany}
\author{C. Rockstuhl}
\affiliation{Institute of Condensed Matter Theory and Solid State Optics, Abbe
Center of Photonics, Friedrich-Schiller-Universit\"{a}t Jena, D-07743
Jena, Germany}

\begin{abstract}
An optical nanoantenna and adjacent atomic systems are strongly coupled
when an excitation is repeatedly exchanged between these subsystems
prior to its eventual dissipation into the environment. It remains
challenging to reach the strong coupling regime but it is equally
rewarding. Once being achieved, promising applications as signal processing
at the nanoscale and at the single photon level would immediately
come into reach. Here, we study such hybrid configuration from different
perspectives. The configuration we consider consists of two identical atomic systems, described
in a two-level approximation, which are strongly coupled to an optical nanoantenna.
First, we investigate when this hybrid system requires a fully quantum
description and provide a simple analytical criterion. Second, a design
for a nanoantenna is presented that enables the strong coupling regime.
Besides a vivid time evolution, the strong coupling is documented
in experimentally accessible quantities, such as the extinction spectra.
The latter are shown to be strongly modified if the hybrid system
is weakly driven and operates in the quantum regime. We find that
the extinction spectra depend sensitively on the number of atomic
systems coupled to the nanoantenna.
\end{abstract}


\maketitle

\section{Introduction}

Metallic optical nanoantennas have proven perfect in tailoring light-matter
interactions at the nanoscale. They allow to drastically change the
spontaneous emission rates of adjacent atomic systems or their radiation
properties, see e.g. Refs.~\cite{Anger2006,Giessen2010,vanHulst2010,Zhu2012,Filter2013Gr,Mohtashami2013}.
Although the modified light-matter interaction manifests in a multitude
of phenomena, all of them are eventually promoted by the same underlying
principle, that metallic optical nanoantennas can support strongly
localized surface plasmon polaritons. This is at the heart of all
observations and entails that their coupling to far-field radiation
and quantum systems can be engineered on purpose \cite{novotny2011antennas,Rogobete2007}.

Recent advances in nanotechnology eventually permitted to fabricate
nanoantennas with a precision down to the atomic scale \cite{Hecht2012}.
This implies that a precise arrangement of e.g. quantum dots, molecules,
or atoms close to a carefully designed nanoantenna is feasible. It
has been already shown that in such situations remarkable new phenomena
can be expected where the huge enhancement of dipole-forbidden transitions
in the gap of a dimer nanoantenna may serve as a representative example
\cite{Kern2012,Filter2012}. The tremendous spatial localization of
the plasmonic mode permits a strong coupling of quantum systems to
nanoantennas. The strong coupling regime is characterized by a transition
from irreversible spontaneous emission and nonradiative damping processes
to a reversible energy exchange between nanoantenna and atomic system.
Such a behavior has been reported for cavities operating in the infrared
and visible spectral domain \cite{Yoshie2004,Reithmaier2004,Aoki2006}.
To achieve
strong coupling is of paramount importance with respect to applications
in deterministic quantum computation and for high-power emission of
nonclassical light into predefined directions.
\begin{figure}
\begin{centering}
\includegraphics[width=8.6cm,keepaspectratio]{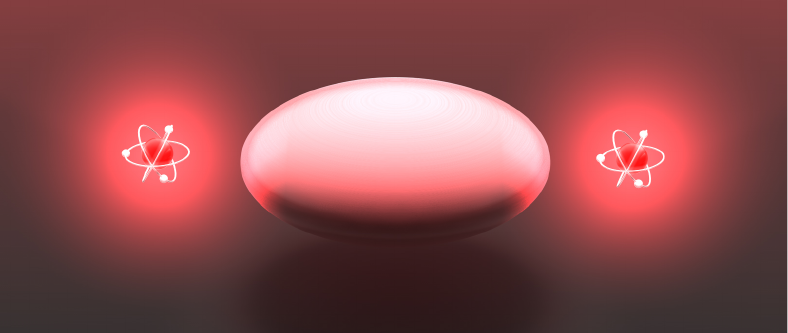}
\par\end{centering}

\caption{\label{fig:coupled_tls} A general scheme of the considered hybrid
system. A nanoantenna is strongly coupled to two atoms and excited
by an external driving field.}
\end{figure}

In our contribution we want to go one step further and consider two
atoms, rather than a single, isolated one, strongly coupled to a nanoantenna.
The aim is to demonstrate that in this case strong coupling effects
appear\textbf{ }more pronounced where one purpose of the nanoantenna
is to strongly increase the interaction between both atoms. Moreover,
the structure is highly interesting since a much larger splitting
of the energy levels of the hybrid system may be anticipated when
compared to that of bare atoms isolated from the nanoantenna. Perspectively,
this may suggest an alternative route towards artificial atoms with
engineered energy levels. Furthermore, the properties of the hybrid
system are shown to sensitively depend on the number of atoms or molecules
involved, paving the way for ultra-sensitive devices operating on
the single molecule level.

Moreover, besides being of importance from an applied perspective,
the setup is essential for basic science since it constitutes a system
of rich dynamics that can be operated in different regimes where each
regime requires a well-adapted approach to fully grasp its properties.
Specifically, the nanoantennas themselves may be described at different
levels of approximation.

The simplest approach is to consider the nanoantenna as a passive
system which can significantly influence the atoms' dynamics whereas
the properties of the nanoantenna remain unaffected.
If such approximation holds, the nanoantenna
is treated as a classical harmonic oscillator as frequently done in
the literature \cite{Yan2008,Artuso2011}. This simplifies the treatment
considerably but prevents the observation of effects associated with
the quantum nature of the nanoantenna like Rabi splitting or anti-bunching
of its emitted light \cite{Meystre1999}.

On the contrary, one may account for the full dynamics of the electrons
inside the nanoantenna. Such an exhaustive treatment is required if
the atoms are placed only a few angstroms off the nanoantenna such
that electron-spill-out and quantum tunneling effects become relevant
\cite{Zuloaga2010,Manjavacas2011}. However, these effects have not
to be considered in most experimentally accessible situations.

In-between, however, lies a regime where the nanoantenna can be considered
as a harmonic oscillator but which requires a proper quantization
\cite{Hohenester2008,Waks2010,Dzsotjan2011,GovorovQuantumFano,Hummer2013}.
The approach permits the description of the rich quantum behavior
of the hybrid system. It is especially useful to study effects at
low power levels where only a few photons are involved. However, it
is \textit{a priori} not clear whether such elaborated approach is
necessary or whether the semiclassical treatment is already sufficient.

Therefore, at first we will develop both a semiclassical and a fully
quantum theory for the problem where an atom described in a two-level
approximation is coupled to an optical nanoantenna. In Section \ref{sec:comparison} we
are going to compare the results of both approaches and we will derive
an unambiguous criterion that can be used to decide, on the base of experimentally
accessible quantities, which of the two approaches is necessary. Second,
in Section \ref{sec:design} we will explicitly discuss the design
of a nanoantenna that can be operated in the strong coupling regime.
After that, in Section \ref{sec:absorption} we are going to study
the impact of the strong coupling on the extinction spectra of a hybrid
system where two atoms are coupled to the optical nanoantenna. We
will find that the presence of the atoms strongly affects both absorption
and scattering properties of the hybrid system. After concluding on
our findings we will provide in elaborated appendices details of our
calculations and results that support our conclusions from the main
body of the manuscript. In App.~\ref{app:class_field_dynamics} the
equations of motion for the semiclassical formulation will be derived.
Appendix~\ref{app:coupling_constants} details our method to calculate
the coupling constants of the nanoantenna to the atoms and to the
driving field from numerical simulations. Also, the estimation for
the radiative and nonradiative decay rates of the nanoantenna will
be given. In App.~\ref{app:eigenstates}, in the framework of the
fully quantum approach the eigenstates and -energies of the hybrid
system are evaluated in detail.

\section{Model\label{sec:model}}

We consider two identical two-level systems. We refer to them as \textit{atoms},
but they could equally describe molecules, quantum dots, NV centres
in diamond, etc.\cite{Benson2012}. The two-level-systems shall be
symmetrically placed next to a mirror-symmetric nanoantenna that is
excited with a driving field propagating along the symmetry axis.
Schematically, the situation is shown in Fig.~\ref{fig:coupled_tls}.
This high symmetry configuration has been chosen for the sake of a
reasonable simplicity in our treatment but does not constitute any
limitation.

As discussed above, the theoretical description of such a system may
be performed in several approximations. In subsection \ref{sub:fully_quantum}
we consider a fully quantum model including a quantum description
of the nanoantenna itself. In practice, such approach requires a considerable
numerical effort, unless the excitation field is rather weak and the
total system remains approximately at the single excitation level.
A considerable simplification and limitation of numerical efforts
is provided by a mean-field approximation, where the electromagnetic
field is described classically \cite{Karolina2012}. This semi-classical
treatment will be given in subsection \ref{sub:semiclassical}. Both
models will be compared in the succeeding section.

\subsection{Fully quantum approach\label{sub:fully_quantum}}

In the fully quantum approach, we regard the nanoantenna as a single
mode quantum harmonic oscillator. Then the Hamiltonian in the rotating
frame takes the following form within the rotating wave approximation
\cite{Meystre1999}:
\begin{eqnarray}
H & = & \frac{\hbar\,\Delta\omega_0}{2}\sum_{j=1}^{N_\mathrm{tls}}\left(\sigma_{z}^{(j)}+\unit \right)+\hbar\Delta\omega_\mathrm{na} a^{\dagger}a \label{eq:hamiltonian}\\
 &  & -\hbar\kappa\sum_{j=1}^{N_\mathrm{tls}}\left(\sigma_{+}^{(j)}a+a^{\dagger}\sigma_{-}^{(j)}\right)-
 \hbar\Omega\left(a+a^{\dagger}\right),\nonumber
\end{eqnarray}
with $\Delta\omega_{0}=\omega_{0}-\omega_{\mathrm{dr}}$, $\Delta\omega_{\mathrm{na}}=\omega_{\mathrm{na}}-\omega_{\mathrm{dr}}$.
Here, ${N_{\mathrm{tls}}}=2$ is the number of identical atoms, and
$\omega_{0}$ corresponds to their transition frequency. The operators
$\sigma_{z}^{(j)}=|e^{(j)}\rangle\langle e^{(j)}|-|g^{(j)}\rangle\langle g^{(j)}|$
represent the population inversion in the $j^{\mathrm{th}}$ two-level
system, and $\sigma_{+}^{(j)}=|e^{(j)}\rangle\langle g^{(j)}|$ and
$\sigma_{-}^{(j)}={\sigma_{+}^{(j)}}^{\dagger}$ are the corresponding
creation and annihilation operators of atomic excitation. As usual,
$\{|e^{(j)}\rangle\text{ and }|g^{(j)}\rangle\}$ denote the excited
and ground state of the $j^{\mathrm{th}}$ two-level system. The symbols
$a$ and $a^{\dagger}$ stand for the annihilation and creation operator
of the nanoantenna mode, respectively. Several approximations have been made here.
For simplicity, only the dipolar transitions in the atoms have been taken into account.
In case when such approximation is not valid, a treatment like the one described in
Ref. \cite{Lembessis2013}, should be applied.
Treatment of the nanoantenna as a single-mode harmonic oscillator is also an approximation.
It is assumed that a single resonance dominates the nanoantenna spectrum
around the atomic transition frequency, see also Refs. \cite{Yan2008,Filter2012}
and the discussion in Section \ref{sec:design}.

We note that the single-mode
approach for both the atomic system and the nanoantenna is an approximation
taking only the dipolar transitions in the atomic
system and the nanoantenna into account, see
and App.~\ref{app:coupling_constants}.

The nanoantenna is coherently driven by an external laser beam, assumed
to be monochromatic at frequency $\omega_{\mathrm{dr}}$. The driving
field intensity is related to the Rabi frequency $\Omega$, which
is here taken real for simplicity, see App~\ref{app:coupling_constants}.
The coupling constant between atom and nanoantenna is given by $\kappa$,
identical for both atoms because of their symmetric placement. Note
that $\kappa$ can be assumed constant, if the transition frequency
of the atoms is close to the broad resonance of the nanoantenna of
central frequency $\omega_{\mathrm{na}}$. We neglect the free-space
interaction of the atoms, i.e. the dipole-dipole interaction in the
absence of the nanoantenna, because it is considerably weaker than
the interaction of the atoms due to the nanoantenna. We also neglect
the direct coupling of the driving field to the atoms, as, again,
it is much weaker than the nanoantenna's scattered field at the position
of the atoms.

The dynamics of the hybrid system is described by the Lindblad-Kossakowski
equation \cite{Meystre1999}:
\begin{eqnarray}
\mathrm{i}\hbar\dot{\rho} & = & \left[H,\rho\right]+\mathrm{i}\mathcal{L}_{\mathrm{na}}\left(\rho\right)+\mathrm{i}\mathcal{L}_{\mathrm{tls}}\left(\rho\right),\label{eq:master}
\end{eqnarray}
where $\rho$ is the density operator of the hybrid system and $\mathcal{L}_{\mathrm{na,tls}}\left(\rho\right)$
are Lindblad operators responsible for losses in the nanoantenna and
the atoms, respectively, given by
\begin{eqnarray}
\mathcal{L}_{\mathrm{na}}\left(\rho\right) & = & -\hbar\Gamma\left(a^{\dagger}a\rho+\rho a^{\dagger}a-2a\rho a^{\dagger}\right),\\
\mathcal{L}_{\mathrm{tls}}\left(\rho\right) & = & -\frac{1}{2}\hbar\Gamma_{\mathrm{fs}}\sum_{j=1}^{N_{\mathrm{tls}}}\left(\sigma_{+}^{(j)}\sigma_{-}^{(j)}\rho+\rho\sigma_{+}^{(j)}\sigma_{-}^{(j)}\right.\nonumber \\
 &  & \left.-2\sigma_{-}^{(j)}\rho\sigma_{+}^{(j)}\right)\\
 &  & +\frac{1}{2}\hbar\Gamma_{\mathrm{d}}\sum_{j=1}^{N_{\mathrm{tls}}}\left(\sigma_{z}^{(j)}\rho\sigma_{z}^{(j)}-\rho\right).\nonumber
\end{eqnarray}
In the above expressions $\Gamma=\Gamma_{\mathrm{r}}+\Gamma_{\mathrm{nr}}$
describes radiative and nonradiative losses by the nanoantenna. The
free-space spontaneous emission rate of a single two-level system
is given by $\Gamma_{\mathrm{fs}}$, whereas $\Gamma_{\mathrm{d}}$
is the rate of pure dephasing. An example of the latter would be the
interactions with phonons in quantum dots that affect the coherence
but not the population distribution \cite{Heiss2007}. Typically,
the radiative and nonradiative losses in the metallic nanoparticle
are much stronger than all losses in the atoms.

\subsection{Semiclassical approach\label{sub:semiclassical}}

In the preceding subsection we have formally introduced a fully quantum
description of the atoms coupled to a nanoantenna. Now the same physical
situation will be considered but the description of the nanoantenna
will be approximated by a classical equation of motion. The state
of the atomic system may then be described by its density operator
which we denote as $\rho^{\mathrm{sc}}$. Its evolution follows the
Lindblad-Kossakowski equation
\begin{equation}
\mathrm{i}\hbar\dot{\rho}^{\mathrm{sc}}=\left[H^{\mathrm{sc}},\rho^{\mathrm{sc}}\right]+\mathrm{i}\mathcal{L}_{\mathrm{tls}}\left(\rho^{\mathrm{sc}}\right),\label{eq:simclassical_master_equation_2TLS}
\end{equation}
with the semiclassical Hamiltonian \cite{Meystre1999}
\begin{eqnarray}
H^\mathrm{sc} & = & \frac{\hbar\Delta\omega_0}{2}\sum_{j=1}^{N_{\mathrm{tls}}}\left(\sigma_{z}^{(j)}+\unit \right)\label{eq:semiclassical_Hamiltonian}\\
& & -\hbar\kappa\sum_{j=1}^{N_{\mathrm{tls}}}\left[\sigma_{+}^{(j)}\alpha\left(t\right)+
\alpha^{\star}\left(t\right)\sigma_{-}^{(j)}\right]\ . \nonumber
\end{eqnarray}
For the sake of comparison to the fully quantum Hamiltonian $H$,
we denote the Rabi frequency of the scattered field by $\kappa\alpha\left(t\right)$,
with the classical dimensionless amplitude \mbox{$\alpha\left(t\right)\in\mathbb{C}$}.

The time-dependent dipole moment of each atom acts naturally as electrodynamic
source. Thus, the overall field is a superposition of contributions
from the atoms and the nanoantenna. Usually the field generated by
the $j^{\mathrm{th}}$ atom is expressed by its mean transition dipole
moment, $\mathbf{E}\left(\mathbf{r},t\right)\propto\langle\mathbf{d}^{(j)}\rangle=\mathbf{d}_{\mathrm{ge}}^{\mathrm{(j)}}{\rho_{\mathrm{eg}}^{\mathrm{sc}}}^{(j)}$
\cite{Yan2008}, where ${\rho_{\mathrm{mn}}^{\mathrm{sc}}}^{(j)}={{\rho_{\mathrm{nm}}^{\mathrm{sc}}}^{(j)}}^{\star}=\langle m|{\rho^{\mathrm{sc}}}^{(j)}|n\rangle$
is an element of the reduced density matrix of the $j^{\mathrm{th}}$
atom, and the asterisk stands for the complex-conjugate. Similarly,
$\mathbf{d}_{\mathrm{ge}}^{\mathrm{(j)}}=\langle g^{(j)}|\mathbf{d}^{\mathrm{(j)}}|e^{(j)}\rangle$
corresponds the matrix element of the $j^{\mathrm{th}}$ atom dipole
moment operator. As derived in App.~\ref{app:class_field_dynamics},
the evolution equation of the field in the slowly-varying envelope
approximation $|\dot{\alpha}|\ll|\alpha\omega_{\mathrm{dr}}|$ and
for $\Gamma\ll\omega_{\mathrm{dr}}$ is given by
\begin{eqnarray}
\dot{\alpha}\left(t\right) & = & -\left(\Gamma+i\Delta\omega_{\mathrm{na}}\right)\alpha\left(t\right)\label{eq:alpha_evolution_mean}\\
 &  & +i\left[\kappa\sum_{j}{\rho_{eg}^{\mathrm{sc}}}^{(j)}(t)+\Omega\right].\nonumber
\end{eqnarray}

Such a description is an approximation which turns out to be valid
just for very weak excitations as we will show in the following section.

\section{Comparison of semiclassical and fully quantum approaches\label{sec:comparison}}

One of the main issues addressed in this paper concerns the identification
of conditions where the semiclassical and the fully quantum approaches
yield equivalent results. To this end, we will consider the simplest
scenario where a single atom is coupled to the nanoantenna at first.
We will directly compare the evolution equations of the atomic operators
in the fully quantum approach, obtained in the Heisenberg picture,
with the corresponding ones of the atomic density matrix and of the
field amplitude in the semiclassical approach. A limit will be identified
where both approaches agree to a good approximation. An interpretation
in terms of correlation functions will also be provided.

As we will show, the very source of discrepancies between both approaches
is the interaction term proportional to $\kappa$, which we will now
focus on. For this reason we consider for a while the simplified lossless
case, and also set $\Omega=0$, but assume that the excitation is
initially present in the coupled system. For instance, the atom is
initially excited and/or photons are present in the field of the nanoantenna.

Directly from the Heisenberg equation $\dot{A}=-i/\hbar[A,H]+\frac{\partial A}{\partial t}$,
which describes the evolution of any operator $A$, we obtain the
evolution of the field annihilation operator in the rotating frame
in the fully quantum picture:
\begin{equation}
\dot{a}\left(t\right)=-i\Delta\omega_{\mathrm{na}}a\left(t\right)+i\kappa\sigma_{-}\left(t\right),\label{eq:a_evolution}
\end{equation}
and of the atomic operators:
\begin{eqnarray}
\dot{\sigma}_{z}\left(t\right) & = & 2\mathrm{i}\kappa\left[\sigma_{+}\left(t\right)a\left(t\right)-a^{\dagger}\left(t\right)\sigma_{-}\left(t\right)\right],\label{eq:sigma_evolution}\\
\dot{\sigma}_{-}\left(t\right) & = & -\mathrm{i}\Delta\omega_{0}\sigma_{-}\left(t\right)-\mathrm{i}\kappa\sigma_{z}\left(t\right)a\left(t\right).\label{eq:sigma_evolutionII}
\end{eqnarray}
The first equation can be formally integrated to give
\begin{eqnarray}
{a}\left(t\right) & = & a\left(0\right)e^{-i\Delta\omega_{\mathrm{na}}t}\label{eq:a_lossless}\\
 &  & +i\int_{0}^{t}\kappa\sigma_{-}\left(t^{\prime}\right)e^{-i\Delta\omega_{\mathrm{na}}\left(t-t^{\prime}\right)}dt^{\prime}.\nonumber
\end{eqnarray}
Inserting this resut into equations (\ref{eq:sigma_evolution}-\ref{eq:sigma_evolutionII})
leads to the following evolution equations for the expectation values:
\begin{eqnarray}
\langle\dot{\sigma}_{z}\left(t\right)\rangle & = & 2\mathrm{i}\kappa\left[\langle\sigma_{+}\left(t\right)a\left(0\right)\rangle e^{-i\Delta\omega_{\mathrm{na}}t}-\mathrm{c.c.}\right]\\
 &  & -2\kappa^{2}\int_{0}^{t}\langle\sigma_{+}\left(t\right)\sigma_{-}\left(t^{\prime}\right)\rangle e^{-i\Delta\omega_{\mathrm{na}}\left(t-t^{\prime}\right)}dt^{\prime}\nonumber \\
 &  & -2\kappa^{2}\int_{0}^{t}\langle\sigma_{+}\left(t^{\prime}\right)\sigma_{-}\left(t\right)\rangle e^{i\Delta\omega_{\mathrm{na}}\left(t-t^{\prime}\right)}dt^{\prime}\nonumber \\
\langle\dot{\sigma}_{-}\left(t\right)\rangle & = & -\mathrm{i}\Delta\omega_{0}\langle\sigma_{-}\left(t\right)\rangle\\
 &  & -\mathrm{i}\kappa\langle\sigma_{z}\left(t\right)a\left(0\right)\rangle e^{-i\Delta\omega_{\mathrm{na}}t}\nonumber \\
 &  & +\kappa^{2}\int_{0}^{t}\langle\sigma_{z}\left(t\right)\sigma_{-}\left(t^{\prime}\right)\rangle e^{-i\Delta\omega_{\mathrm{na}}\left(t-t^{\prime}\right)}dt^{\prime},\nonumber
\end{eqnarray}
where $\mathrm{c.c.}$ stands for the complex-conjugate. We consider
here the evolution of expectation values of the atomic operators because
it can be directly compared with the evolution of the corresponding
elements of the density matrix, i.e. $\langle\sigma_{z}(t)\rangle=\rho_{ee}^{\mathrm{sc}}(t)-\rho_{gg}^{\mathrm{sc}}(t)$,
$\langle\sigma_{-}(t)\rangle=\rho_{eg}^{\mathrm{sc}}(t)$.

Similarly, in the semiclassical description we integrate equation
(\ref{eq:alpha_evolution_mean}) with $\Gamma$ and $\Omega$ set
to zero. Next, we insert it into the Lindblad-Kossakowski equation
(\ref{eq:simclassical_master_equation_2TLS}) with $\Omega=0$ and
$\Gamma_{\mathrm{fs}}=\Gamma_{\mathrm{d}}=0$, to arrive at:
\begin{eqnarray}
\dot{\rho}_{ee}^{\mathrm{sc}}-\dot{\rho}_{gg}^{\mathrm{sc}} & = & 2\mathrm{i}\kappa\left[\rho_{ge}^{\mathrm{sc}}\left(t\right)\alpha\left(t\right)-\alpha^{\star}\left(t\right)\rho_{eg}^{\mathrm{sc}}\left(t\right)\right]\label{eq:Bloch_eqs}\\
 & = & 2\mathrm{i}\kappa\left[\rho_{ge}^{\mathrm{sc}}\left(t\right)\alpha\left(0\right)e^{-\mathrm{i}\Delta\omega_{\mathrm{na}}t}-\mathrm{c.c.}\right]\nonumber \\
 &  & -2\kappa^{2}\int_{0}^{t}\rho_{ge}^{\mathrm{sc}}\left(t\right)\rho_{eg}^{\mathrm{sc}}\left(t^{\prime}\right)e^{-i\Delta\omega_{\mathrm{na}}\left(t-t^{\prime}\right)}dt^{\prime}\nonumber \\
 &  & -2\kappa^{2}\int_{0}^{t}\rho_{ge}^{\mathrm{sc}}\left(t^{\prime}\right)\rho_{eg}^{\mathrm{sc}}\left(t\right)e^{i\Delta\omega_{\mathrm{na}}\left(t-t^{\prime}\right)}dt^{\prime}\nonumber \\
\dot{\rho}_{eg}^{\mathrm{sc}} & = & -\mathrm{i}\Delta\omega_{0}{\rho}_{eg}^{\mathrm{sc}}-\mathrm{i}\kappa\alpha\left(t\right)\left[\rho_{ee}^{\mathrm{sc}}\left(t\right)-\rho_{gg}^{\mathrm{sc}}\left(t\right)\right]\\
 & = & -\mathrm{i}\Delta\omega_{0}{\rho}_{eg}^{\mathrm{sc}}\nonumber \\
 &  & -\mathrm{i}\kappa\left[\rho_{ee}^{\mathrm{sc}}\left(t\right)-\rho_{gg}^{\mathrm{sc}}\left(t\right)\right]\alpha\left(0\right)e^{-i\Delta\omega_{\mathrm{na}}t}\nonumber \\
 &  & +\kappa^{2}\int_{0}^{t}\left[\rho_{ee}^{\mathrm{sc}}\left(t\right)-\rho_{gg}^{\mathrm{sc}}\left(t\right)\right]\rho_{eg}^{\mathrm{sc}}\left(t^{\prime}\right)\times\nonumber \\
 &  & \times e^{-i\Delta\omega_{\mathrm{na}}\left(t-t^{\prime}\right)}dt^{\prime}.\nonumber
\end{eqnarray}

Now by directly comparing the equations obtained in both descriptions
we note that the semiclassical approach leads to nonlinear terms of
the type $\rho_{ij}^{\mathrm{sc}}(t)\rho_{kl}^{\mathrm{sc}}(t^{\prime})$
or, equivalently, $\langle\sigma_{p}(t)\rangle\langle\sigma_{q}(t^{\prime})\rangle$,
with $i,j,k,l=e,g$ and $p,q=z,+,-$. On the other hand, from the
analysis of the Heisenberg equations of motion, i.e. without the mean
field approximation, we can see that terms such as $\langle\sigma_{p}(t)\sigma_{q}(t^{\prime})\rangle$
appear in the equations of motion instead \cite{Chen2012coherent}.

Both results are only equivalent if the atomic operators are uncorrelated,
i.e. if $\langle\sigma_{p}(t)\sigma_{q}(t^{\prime})\rangle\approx\langle\sigma_{p}(t)\rangle\langle\sigma_{q}(t^{\prime})\rangle$.
A similar problem has been investigated in Ref.~\cite{Dorner2002},
where the authors use a harmonic oscillator model for a two-level
system and demonstrate that the condition of uncorrelated system operators
is fulfilled for a harmonic oscillator initially it its ground state.
A harmonic oscillator is indeed a good model of a two-level system
if its first excited state occupation probability is small, and the
doubly and higher excited states are not relevant.

Likewise, a two-level system is a good model for a harmonic oscillator
if the system is approximately in its ground state. Then the bosonic
commutation rule can be recovered for the annihilation and creation
operators $\left[\sigma_{-},\sigma_{+}\right]=-\sigma_{z}\approx1$
and one may apply the result of Ref.~\cite{Dorner2002} to the case
of a two-level system. Moreover, only in such case the two-level system
is a source of coherent light, which can be accurately described by
the semiclassical approximation. Only if the condition of uncorrelated
system operators holds true, i.e. the two-level system has to stay
approximately in its ground state throughout the entire evolution,
the mean field approximation, and so the semiclassical approach, are
valid \cite{Waks2010}. It is important to note that the applicability
of this restriction does not depend on the coupling strength. Furthermore,
our condition holds in general for any situation where atomic systems
interact with light rather than only for coupling of two-level systems
to nanoantennas considered here.

For simulations it is often crucial to find a strict criterion for
the validity of the semiclassical approximation. Such a criterion
can be found by deriving the steady-state solution of equations (\ref{eq:Bloch_eqs}),
where the driving field or loss rates are no longer assumed to be
zero. The assumption that must be fulfilled for the semiclassical
approximation to be valid is that the excited state occupation of
either atomic system is small, i.e. $1\approx\rho_{gg}^{\mathrm{sc}}\gg\rho_{ee}^{\mathrm{sc}}$.
Then, we find
\begin{eqnarray}
\rho_{ee}^{\mathrm{sc}} & \approx & \frac{2\Gamma_{\mathrm{dec}}}{\Gamma_{\mathrm{fs}}}\frac{\kappa^{2}\Omega^{2}}{D}\ll1,\\
D & = & (\Gamma_{\mathrm{dec}}^{2}+\Delta\omega_{0}^{2})(\Gamma^{2}+\Delta\omega_{\mathrm{na}}^{2})\nonumber \\
 &  & +2(\Gamma_{\mathrm{dec}}\Gamma-\Delta\omega_{0}\Delta\omega_{\mathrm{na}})\kappa^{2}+\kappa^{4},\nonumber
\end{eqnarray}
where $\Gamma_{\mathrm{dec}}=\frac{1}{2}\Gamma_{\mathrm{fs}}+\Gamma_{\mathrm{d}}$
is the total decoherence rate of an atom, which includes contributions
from spontaneous emission and pure dephasing processes. In the resonant
case with $\omega_{\mathrm{dr}}=\omega_{\mathrm{na}}=\omega_{0}$,
the validity criterion for the semiclassical approximation simplifies
to
\begin{eqnarray}
\rho_{ee}^{\mathrm{sc}} & \approx & \frac{2\Gamma_{\mathrm{dec}}}{\Gamma_{\mathrm{fs}}}\frac{\kappa^{2}\Omega^{2}}{(\Gamma_{\mathrm{dec}}\Gamma+\kappa^{2})^{2}}\ll1.\label{eq:semiclassical_criterion_resonant}
\end{eqnarray}
This condition of weak driving fields is confirmed in numerical simulations
using Mathematica 7 \cite{wolfram1999mathematica}. The calculations
were carried out by numerically solving the Lindblad-Kossakowski equation
(\ref{eq:master}) in the fully quantum approach. For the semiclassical
approach, Eqs.~(\ref{eq:simclassical_master_equation_2TLS}) and
(\ref{eq:alpha_evolution_mean}) were numerically solved, respectively.

\begin{center}
\begin{figure}[t]
\includegraphics[width=8cm]{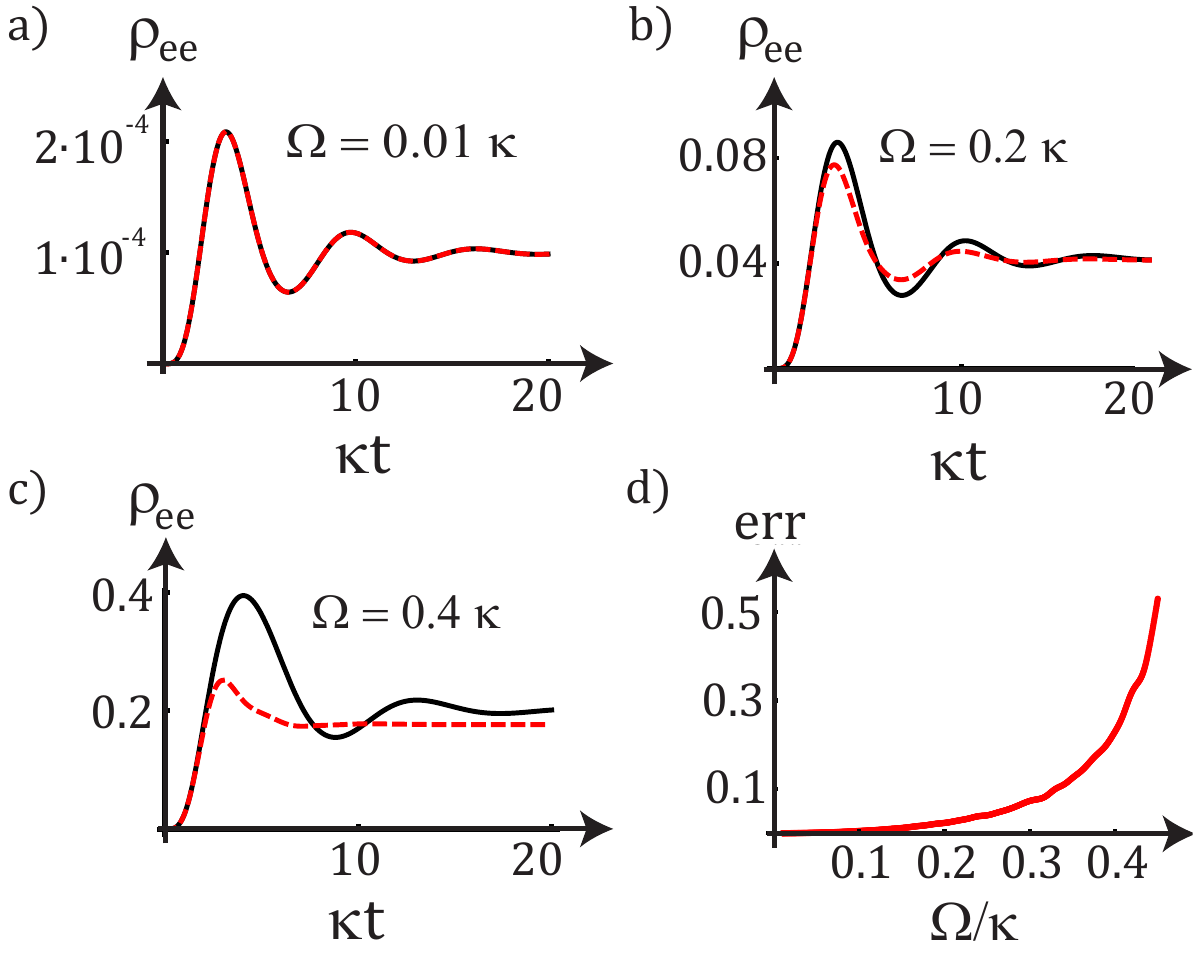} \begin{caption} {Excited state
occupation probability of the atom vs. normalized time for different
Rabi frequencies $\Omega$ of the driving field and $\omega_{\mathrm{dr}}=\omega_{\mathrm{na}}=\omega_{0}$,
$\Gamma=0.5\kappa$, $\Gamma_{\mathrm{fs}}=10^{-9}\kappa$, $\Gamma_{\mathrm{d}}=0$.
The black solid line corresponds to the semiclassical and the red
dashed line to the fully quantum results. (a) For weak driving fields
a perfect agreement is obtained. (b, c) For stronger driving fields
both the time evolution and the steady state results are incorrectly
evaluated in the semiclassical approximation (\ref{eq:semiclassical_criterion_resonant}).
(d) The relative error of the steady-state results grows fast with
increasing $\Omega/\kappa$ ratio.\label{semiclassical}} \end{caption}
\end{figure}

\par\end{center}

We have performed simulations for different driving strengths assuming
$\omega_{\mathrm{dr}}=\omega_{\mathrm{na}}=\omega_{0}$, $\Gamma=0.5\kappa$,
$\Gamma_{\mathrm{fs}}=10^{-9}\kappa$, $\Gamma_{\mathrm{d}}=0$. Initially,
the atom is assumed to be in its ground state and the field amplitude
of the nanoantenna vanishes: $\rho(t=0)=|g\rangle\langle g|\otimes|0\rangle\langle0|$
is the initial condition for the fully quantum case, whereas $\rho^{\mathrm{sc}}(t=0)=|g\rangle\langle g|$
and $\alpha(t=0)=0$ correspond to that of the semiclassical one.
In the fully quantum approach we perform the calculations in a Hilbert
space truncated at sufficiently high number $k_{\mathrm{max}}$ of
excitations in the nanoantenna. Here it suffices to set $k_{\mathrm{max}}=10$.
In Fig.~\ref{semiclassical}, the excited state occupation probability
of the atom is compared when calculated in the semiclassical (black
solid line) and the fully quantum approach (red dashed line), for
increasing intensities of the driving field {[}panels (a)-(c){]}.
For weak fields, and thus small excited state occupations, the results
are in perfect agreement (a). Slightly stronger fields result in discrepancies
in time evolution, but the steady state occupation coincides in both
approaches (b). This is no longer the case for rather strong driving
fields that result in considerable excitation probabilities of the
atom (c). In panel (d) the steady-state relative error, defined as
$\mathrm{err}\equiv\left.\left|\rho_{\mathrm{ee}}^{\mathrm{sc}}-\rho_{\mathrm{ee}}\right|/\rho_{\mathrm{ee}}\right|_{t\rightarrow\infty}$,
is shown to grow fast with the ratio of the Rabi frequency of the
driving field $\Omega$ to the coupling constant $\kappa$.

On more analytical grounds, we can first examine the simplest case
where the spontaneous emission is the only source of decoherence,
i.e. $\Gamma_{\mathrm{d}}=0$ and $\Gamma_{\mathrm{dec}}=\frac{1}{2}\Gamma_{\mathrm{fs}}$.
Then, the prefactor in Eq.~(\ref{eq:semiclassical_criterion_resonant})
is unity because of \mbox{$2\Gamma_{\mathrm{dec}}/{\Gamma_{\mathrm{fs}}}=1$.}
Moreover, for small losses Eq.~(\ref{eq:semiclassical_criterion_resonant})
may be further simplified. Then we find a condition for the validity
of the semiclassical approach as $\Omega\ll\kappa$. If that condition
holds $\rho_{ee}^{\mathrm{sc}}\ll1$, i.e., the atom will approximately
remain in its ground state. If losses dominate, i.e. for atoms weakly
coupled to a nanoantenna, the semiclassical approximation can be applied
as long as $\Omega\ll\Gamma\Gamma_{\mathrm{fs}}/\kappa$.

If dephasing is additionally present $(\Gamma_{\mathrm{d}}\neq0)$,
the excited state occupation is always increased (not shown) with
respect to the case of pure spontaneous emission, which makes condition
(\ref{eq:semiclassical_criterion_resonant}) more difficult to be
fulfilled. For small driving field intensities the peak value of the
excited-state occupation probability is equal to $\rho_{\mathrm{ee,max}}^{\mathrm{sc}}\approx\Omega^{2}/2\Gamma_{\mathrm{fs}}\Gamma$
and is reached for $\Gamma_{\mathrm{dec}}=\kappa^{2}/\Gamma$. Thus
we arrive at a strong worst-case-scenario criterion valid for an arbitrary
dephasing rate and an arbitrary coupling strength: $\Omega^{2}\ll\Gamma_{\mathrm{fs}}\Gamma$.

This result may seem counterintuitive. One might have expected that
the semiclassical approximation is valid in the limit of sufficiently
strong rather than weak fields. This result can be understood as follows.
For weak driving fields, in good approximation the atoms are in their
ground states. Then, they behave as harmonic oscillators which can
be accurately described by the semiclassical approach which is in
accordance with the results presented in Ref.~\cite{Waks2010}. For
stronger fields, the approximation of the two-level systems as harmonic
oscillators breaks down and consequently the semiclassical approach
too. However, for even stronger fields the feedback from the atoms
is only of minor relevance. Mathematically, this interaction region
can be defined by $|\alpha|^{2}\gg1$ which naturally leads to $\Omega\gg\Gamma\text{ ,}\kappa$.

\section{Design of the nanoantenna\label{sec:design}}

In this section we aim at designing a nanoantenna that allows achieving
the strong coupling regime. The latter can be defined by \cite{wallraff2004strong}
\begin{equation}
\kappa>\Gamma.\label{eq:condition_strong_coupling}
\end{equation}
This condition suggests that an excitation is exchanged between the
atomic and the nanoantenna subsystems prior to its eventual dissipation
into the environment. We assume the decay and decoherence rates in
the atoms small in comparison with the losses by the nanoparticle.
Slightly differently to Eq.~(\ref{eq:condition_strong_coupling})
strong coupling might also be defined with regard to the emergence
of a dressed state, see \cite{Andreani1999,savasta2010nanopolaritons}.

On our path to a design that allows for the strong coupling regime,
we are going to investigate how coupling constants and loss rates
can be tailored by varying shape and size of the nanoantenna. As a
main result of this section we shall find that strong coupling can
be achieved if the characteristic spatial dimensions of the nanoantenna
are small. The size of the nanoantenna should not be larger than a
few tens of nanometers and the separation of the elements forming
the nanoantenna should be of the order of a few nanometers. Small
spatial dimensions are eventually the crucial condition since they
guarantee sufficiently small mode volumes as required to reach the
strong coupling regime. In the following we analyze the nanoantennas
in terms of two generic parameters that determine their radiative
properties and are frequently exploited while engineering nanoantennas:
the nanoantenna efficiency $\eta\equiv\Gamma_{\mathrm{r}}/\Gamma$
and the Purcell factor $F$. The efficiency is a measure for the fraction
of radiative energy loss $\Gamma_{\mathrm{r}}$ by the nanoantenna
when compared to its total energy loss $\Gamma=\Gamma_{\mathrm{r}}+\Gamma_{\mathrm{nr}}$
\cite{NovotnyHecht}. As discussed earlier, the coupling to higher
order modes was neglected so far. However, not to artificially overestimate
the nanoantenna's efficiency, we calculate $\eta$ for a nanoantenna
excited by dipoles at the positions of the atoms.

The Purcell factor is here understood as a measure for the nanoantenna's
capability to enhance the radiation of a dipole source \cite{Bharadwaj2009}.
It naturally depends on the position of the source with respect to
the nanoantenna. We find the Purcell factor as the ratio
of the total energy flux calculated with and without nanoantenna.
The latter case corresponds to the free-space emission rate $\Gamma_{\mathrm{fs}}$.
Note that the Purcell factor introduced here is different from that
used in the context of cavity QED, i.e. the enhancement of the decay
rate of an atom \cite{Meystre1999,NovotnyHecht}. The decay rate of
atoms interacting with a nanoantenna can be enhanced by radiative
and/or nonradiative loss channels of the nanoantenna. Both losses
coincide only for nanoantennas with the efficiency equal to $1$. Our
results prove that there is a trade-off between the nanoantenna efficiency
(or the Purcell factor) and the coupling strength that can be achieved
with a nanoantenna.

Our electromagnetic simulations were performed
with COMSOL MULTIPHYSICS simulation platform where the dispersive
permittivity has been fully considered \cite{palik}.
The methods that we use to compute the coupling constants and loss rates
for a specific nanoantenna are described in detail in App.~\ref{app:coupling_constants}.
Here, we note that the results were obtained using the plane-wave illumination scheme
for the nanoantenna. In such scheme almost solely the dipolar mode is excited and dominates
the single nanoantenna resonance.
The results might thus change in a different, e.g. dipole, illumination scheme,
where higher-order modes may in general also contribute to the resonance.
However, in case of small nanoparticles, like the ones considered in this paper, this influence can approximately be neglected.
Thus, we do not take these higher order modes into account.

\begin{figure}
\includegraphics[width=8cm]{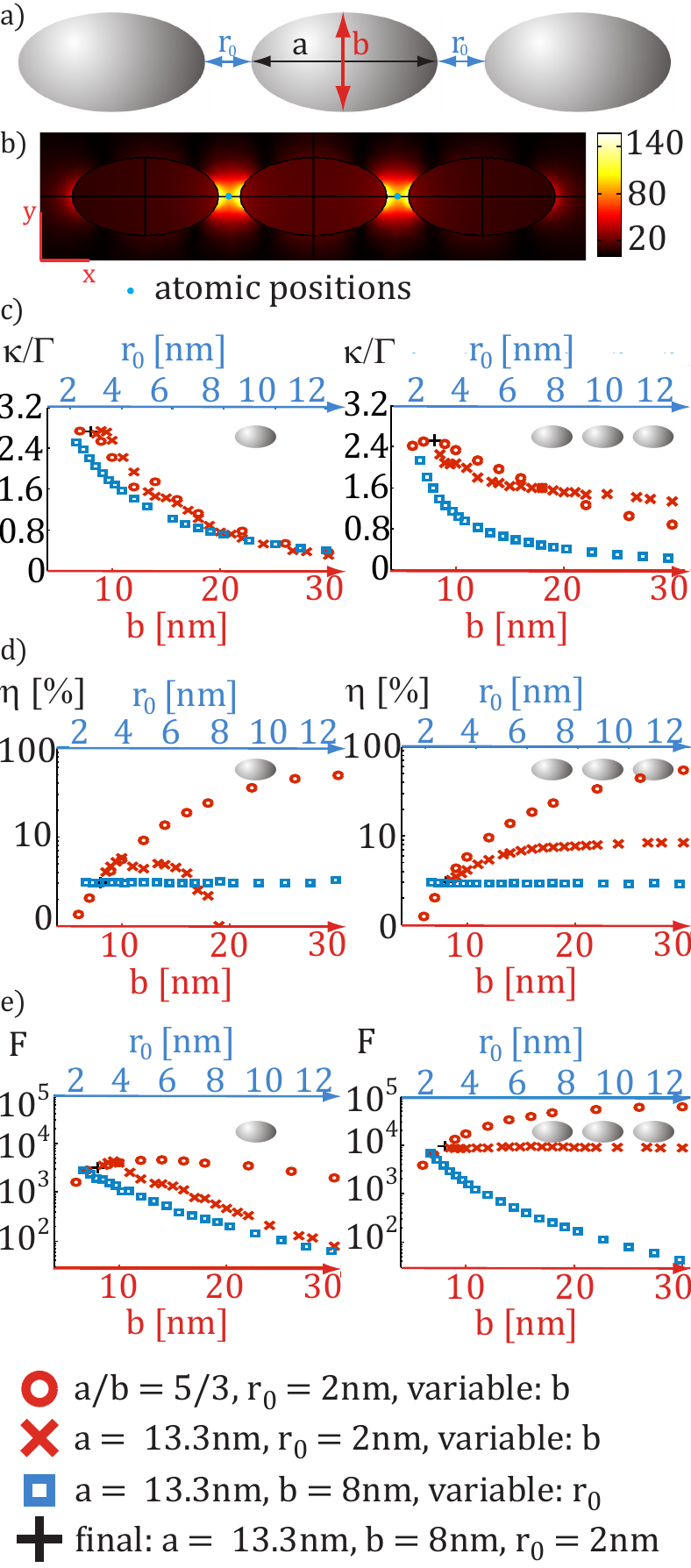} \caption{(a) General scheme of the nanoantenna consisting of three identical
nanospheroids of aspect ratio $a/b$, positioned at distance $r_{0}$
from each other. (b) Distribution of the absolute value of the $x$-polarized
component of the scattered field (normalized by the value of the incoming
field) for three identical nanospheroids with $a=13.3$ nm and $b=8$
nm and separated by $r_{0}=2$ nm. The coupling constant $\kappa$
is proportional to the enhancement of the $x$-polarized scattered
field component at the point where an atom is placed (blue dots). (c) $\kappa/\Gamma$
ratio for $d_{\mathrm{ge}}=6\times10^{-29}$ Cm and various geometrical
parameters of the single- (left) and three-spheroid (right) nanoantenna:
dependence on size
(red circles, constant aspect ratio $a/b=5/3$, $r_{0}=2$ nm), on
aspect ratio (red crosses, constant $a=13.3$ nm, $r_{0}=2$ nm),
and on distance $r_{0}$ (blue squares, $a=13.3$ nm and $b=8$ nm).
The atomic distance from nanoantenna tip is always $r_0/2$.
(d) Nanoantenna
efficiency and (e) Purcell factor, values of parameters as in (c).
The proposed design of a three-spheroid nanoantenna that corresponds
the scattered field distribution of panel (b), is described
in the text and marked in (c)-(e) with the black cross. \label{fig:antenna_parameters}}
\end{figure}

Two basic nanoantenna
geometries which obey the assumptions made in the introduction
are considered from now on: (1) a single or (2) three identical
silver nanospheroids with axis lengths $a$ and $b$.
For prolate nanospheroids $a>b$ holds.
The three-nanospheroid geometry is shown in Fig.~\ref{fig:antenna_parameters}(a).
The nanospheroids are positioned
at a distance $r_{0}$ from each other along the $x$ axis of the
chosen coordinate system with its origin at the center of the middle
nanospheroid.

The reason to compare single- and multiple-nanospheroid geometries
is the following: single nanospheroids are easier to fabricate,
as the experimentally-demanding narrow gap is not required,
and they might be suitable for reaching the strong coupling,
as investigated in \cite{Hohenester2008,Ge2013}.
On the other hand, multiple structures have
the potential to confine and enhance fields much stronger when compared
to isolated nanospheroids. Furthermore, as we will show below,
they are characterized with higher efficiencies and Purcell factors,
which makes them more suitable for applications, e.g., as nonclassical light sources \cite{Mu1992,Maksymov2012}.

To provide a visual impression of the spatial distribution of the
enhanced field, we consider a three-nanospheroid design with the largest $\kappa/\Gamma$
ratio: a nanoantenna made of prolate nanospheroids of $a=13.3$ nm, $b=8$
nm, and $r_{0}=2$ nm, subject to the driving field of frequency
$\omega_{\mathrm{dr}}=3.37\times10^{15}\,\mathrm{s^{-1}}$,
which is the resonance frequency of the nanoantenna. It is worth noting
that an exact resonance with the atomic transition frequency is not
crucial because of the broad resonance of the nanoantenna. The driving
field propagates along the $y$ direction and is polarized in $x$
direction. In Fig.~\ref{fig:antenna_parameters}(b) we show the spatial
distribution of the absolute value of the $x$-polarized component
of the scattered field in the $xy$ plane, normalized to the value
of the incoming field. It is the $x$-component of the enhanced field
which contributes to the coupling constant $\kappa$, as it is parallel
to the assumed direction of the transition dipole moments of the atoms.
Only the scattered, not the total (scattered
+ driving) field is considered here, because the driving field is
considerably weaker and its action on the atom can be neglected to
a good approximation. This is in accordance with the assumptions made
for the Hamiltonians (\ref{eq:hamiltonian},\ref{eq:semiclassical_Hamiltonian}),
considered in Sec. \ref{sec:model}.

For obtaining results displayed in Fig. \ref{fig:antenna_parameters}
we scanned the extinction cross-section \cite{Bohren1998} of every
nanoantenna, subjected to a monochromatic driving field, in the pertinent
frequency domain to find its resonance frequency $\omega_{\mathrm{na}}$.
We assumed a lossless host medium ($\varepsilon=2.2$). Next we considered
a resonantly driven nanoantenna to find both the loss rate $\Gamma$
and the coupling constant $\kappa$ at the positions of the atoms, which
is always taken as $r_0/2$ and in the case of three nanospheroids it is
the point equidistant from two spheroids, as described in App. \ref{app:coupling_constants}.
We set the transition dipole moment of an atom to a rather
high, but realistic value of $d_{\mathrm{ge}}=6\times10^{-29}$ Cm.

In Fig.~\ref{fig:antenna_parameters}(c) the $\kappa/\Gamma$ ratio
for nanoantennas of varying size (red circles) and aspect ratio (red
crosses) is shown. The essential result is that the strong coupling
regime may only be achieved for small nanoantennas (minor axis below
$30$ nm in case of three-, and below $20$ nm in case of single-nanospheroid designs).
Less important is the rather weak dependence on the aspect
ratio $a/b$: the $\kappa/\Gamma$ ratio is larger for prolate objects,
as they can confine and enhance the fields stronger than oblate ones.
Still, the strong coupling can be achieved even for oblate nanospheroids.
However, greater care for the shape must be taken if only one nanospheroid is considered.
In the same figure the dependence on the separation distance $r_{0}$
between the nanospheroids is displayed (blue squares). The distance of the atom
to the nanospheroid is in each case equal to $r_0/2$, so increasing $r_{0}$
means also increasing the distance from the atom to the nanospheroids.
This explains why the field enhancement, and consequently the coupling constant
$\kappa$, drops with distance $r_{0}$. A strong field enhancement,
and thus strong coupling, are obtained if the individual nanospheroids are less
than $4$ nm apart from each other. Interestingly, slightly larger atom-nanoantenna
distances allow for strong coupling in the single-nanospheroid case.

Next we are going to analyze the efficiencies of the above-considered
nanoantennas {[}see Fig.~\ref{fig:antenna_parameters}(d){]}. According
to our investigations, a small size, which is the key for strong coupling,
will result in a poor nanoantenna efficiency,
however, always larger in the multiple-spheroid case.
For nanospheroids of the size at which the strong coupling is reached
($18$ nm in the single- and $30$ nm in the multiple-nanospheroid case)
the efficiency reads $24\%$ and $54\%$, respectively, and drops
dramatically as the size decreases.
In Ref.~\cite{Bohren1998} such scaling of the size-dependent efficiency has been rigorously derived
for spherical particles in terms of scattering and absorption cross
sections.
The efficiency clearly drops for prolate nanoantennas in the muliple-nanospheroid case.
The impact of the adjacent dipoles and of additional nanospheroids
on radiative losses and absorption
turns out to be marginal, so the efficiencies of the proposed nanoantennas
show little dependence on $r_0$.

Relatively higher radiative losses in large nanoantennas result in
an increase of the Purcell factor {[}see Fig.~\ref{fig:antenna_parameters}(d){]}.
Again, it is the size that determines to a large extent the radiative
losses of the nanoparticle.
The Purcell factor is almost constant for a varying aspect ratio
in the case of multiple-spheroid nanoantennas, but it decreases fast
for oblate objects in the case of single spheroids.
Naturally, the Purcell factor drops as the
distance between the nanoantenna and the dipole source grows, since
then their mutual interaction decreases significantly.

We may summarize the above considerations by noting that both single- and
multiple-structure geometries are suitable for strong coupling, although
due to their ability to confine and enhance fields stronger, three-nanospheroid
designs lead to the strong coupling regime at larger nanospheroid sizes.
They are also characterized by larger efficiencies and Purcell factors.
Therefore, the optimal geometry turns out to be the one
always marked with the black cross in Fig. \ref{fig:antenna_parameters}
and used for plotting the field distribution in Fig. \ref{fig:antenna_parameters}(b).
For such design we obtain \mbox{$\Gamma_{\mathrm{nr}}=7.0\times10^{13}$
Hz}, \mbox{$\Gamma_{\mathrm{r}}=6.0\times10^{12}$ Hz}. (Once again,
we note that small nanoantennas designed for strong coupling turn
out to be rather poor emitters: nonradiative losses prevail against
radiative losses by more than one order of magnitude.) The coupling
constant with each of the atoms amounts to $\kappa=2.3\times10^{14}$
Hz. Its large value is responsible for the vivid dynamics of the hybrid
system subject to the driving field, where excitations are exchanged
several times before their eventual dissipation into the environment
via the loss channels of the nanoantenna.

An example of such behavior is presented in Fig.~\ref{fig:shown_strong_coupling},
where the calculations were performed in the fully quantum approach
with the Hilbert space truncated at $k_{\mathrm{max}}=10$. The hybrid
system is initially in its ground state. It is subject to a driving
field of Rabi frequency $\Omega=0.5\kappa$ which quickly leads to
an increase of the probability of a single photon excitation of the
nanoantenna, followed by the probability of a single symmetric excitation
in the atomic subsystem. In the figure the occupation probability
of the state $|S\rangle=\left(|e^{(1)}\rangle|g^{(2)}\rangle+|g^{(1)}\rangle|e^{(2)}\rangle\right)/\sqrt{2}$
is shown. Next also the probability of double excitations rises. For
driving field intensities as small as the one applied here, higher-order
excitations are negligible in the nanoantenna. After the excitations
have flipped several times between the atomic and nanoantenna subsystems,
the hybrid system finally relaxes to a steady state, where an equilibrium
is reached by the driving field and the losses. Note that this result,
with a significant probability of symmetric state occupation, suggests
considerable entangling power of nanoantennas explored e.g. in Refs.
\cite{Gonzalez,Diego}.

The results of this section prove that while engineering a nanoantenna
one has to keep in mind the trade-off between the coupling strength,
achievable with a particular design, and the corresponding efficiency
and ability of the nanoantenna to enhance radiation of dipole emitters.
The size of the structure turns out to be the key parameter, which
needs to be small for achieving the strong coupling regime, where
the hybrid system undergoes a complicated dynamics.

In the following section we will analyze the spectral properties of
the investigated system. For any deviating condition as considered
further below, a suitable nanoantenna that correctly reflects the
situation as considered can be identified out of the data presented
in this section. Figure \ref{fig:antenna_parameters} can serve here
as guideline, from which a possible geometry can be derived for a
realization of a given $\kappa/\Gamma$ rate.

Before we detail the modification of the extinction spectra,
we may comment on the actual experimental feasibility for the suggested
nanoantenna designs. The strong coupling regime can be achieved
with both single- and multiple-nanospheroid geometries.
While the latter generally seem to be more interesting for
applications due to larger values of both efficiencies and Purcell factors,
single nanospheroids may be more feasible from the experimental point
of view, as they do not require very small nanoantenna feed gaps.
In both cases, two main requirements for
an experimental realization can be deduced: a) a highly accurate fabrication
of the nanoantenna itself and b) a precise placement of the atomic
system. Fortunately, both requirements can be accomplished by state-of-the-art
techniques, see e.g. \cite{sidorkin2009sub,Bleuse2011,chen2012single,Alaee2013deep}.

\begin{center}
\begin{figure}[t]
\includegraphics[width=8cm]{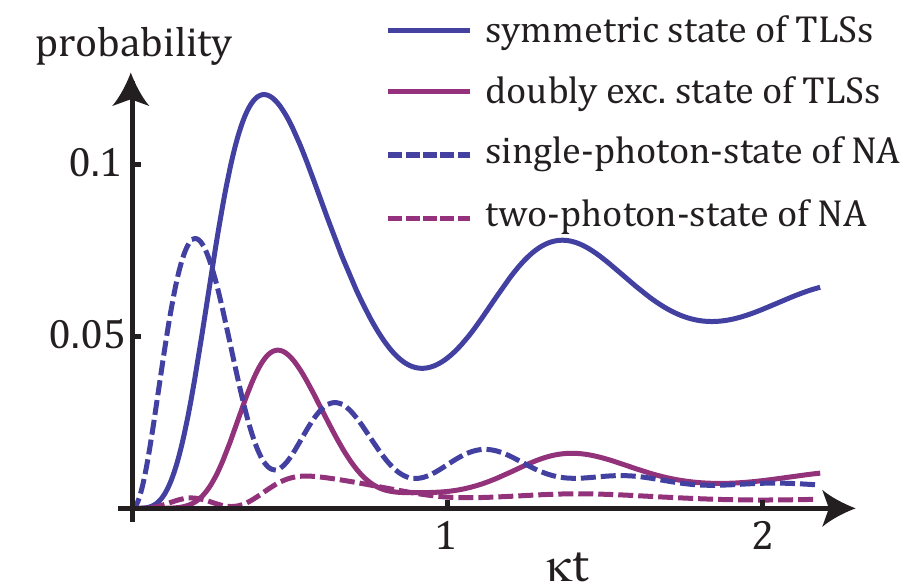} \begin{caption}{Probabilities
of different atomic and nanoantenna excitations vs. normalized time.
Strong coupling between the atomic systems and the nanoantenna is
manifested in the dynamics by the mutual exchange of excitations (atomic
- solid lines, nanoantenna - dashed lines). Peaks in occupation probabilities
of the symmetric (blue solid line, results multiplied by a factor
of $0.5$) and doubly excited (purple solid line) states, correspond
to dips in probabilities of the presence of one (blue dashed line)
and two (purple dashed line) photons in the system. Nanoantenna and
atomic parameters are described in the text.} \label{fig:shown_strong_coupling}
\end{caption}
\end{figure}

\par\end{center}

\section{Modification of Spectra\label{sec:absorption}}

In this section we shall study how the presence of two atoms coupled
to a nanoantenna modifies its extinction spectrum. Such modification
is profound for strong coupling and in the quantum regime. In this
work we may refer to the quantum regime as a situation, where the
mean number of excitations in the system is at the single-quantum
level. We compare the results with the single atom case to prove the
sensitivity of the hybrid quantum system to the number of atomic systems
involved.

The hybrid system permits several loss channels, namely radiative
and non-radiative losses of the nanoantenna and the spontaneous emission
and dephasing in the atomic density matrix. The total extinct
(absorbed and scattered) power is given by
\[
P(\omega_{\mathrm{dr}})\equiv\hbar\omega_{\mathrm{na}}\Gamma\langle a^{\dagger}a\rangle+\sum_{j}^{N_{\mathrm{tls}}}\hbar\omega_{0}\Gamma_{\mathrm{fs}}\langle\sigma_{+}^{\left(j\right)}\sigma_{-}^{\left(j\right)}\rangle\ \ ,
\]
containing the contributions from the nanoantenna and the atoms, respectively.
This quantity is directly accessible in a potential experiment. Here,
$\langle a^{\dagger}a\rangle=\mathrm{Tr}\left[a^{\dagger}a\rho(t\rightarrow\infty)\right]$
denotes the mean number of photons in the system's steady state
$\rho(t\rightarrow\infty)$. Similarly, $\langle\sigma_{+}^{\left(j\right)}\sigma_{-}^{\left(j\right)}\rangle$
corresponds to the excited-state occupation probability of the $j^{\mathrm{th}}$
atom.

In the present study the loss rate of the nanoantenna is much larger
than that of the bare atoms, $\Gamma\gg\Gamma_{\mathrm{fs}}$, see
again Sec.~\ref{sec:design}. Thus, the loss of the nanoantenna can
be used as a an extremely good approximation of the loss of the hybrid
system \cite{TheLukinPaper}.

The spectra are calculated using a freely available quantum optics
toolbox \cite{tan1999computational}. Note that because of the rather
low efficiency of the nanoantenna, i.e. $\Gamma_{\mathrm{nr}}\gg\Gamma_{\mathrm{r}}$,
the extinction spectrum is dominated by the absorption spectrum.

Even though the losses are almost entirely due to the nanoantenna,
the extinction spectrum may be strongly influenced by the two atoms.
In particular, this is the case when the driving field is weak and
the hybrid system remains at the single-excitation level, i.e. in
the quantum regime. This large atomic contribution to the overall
spectrum naturally depends significantly on the coupling as it is
illustrated in Fig.~\ref{fig:spectra_kappa}. For weak coupling {[}see
panel (a){]}, a broad resonance of the nanoantenna dominates the extinction
spectrum and a perturbation at the transition frequency of the atoms
can be observed. For strong coupling {[}panel (b){]}, the plasmonic
and atomic contributions to the spectrum can no longer be distinguished.
The extinction at the atomic transition frequency gets significantly
reduced and Rabi peaks are visible at the sides. Due to the strong
coupling the spectral shift of the extinction peaks can be significantly
exceed the linewidth. Thus strong coupling evokes a large effect of
the atoms on the extinction spectrum of the nanoantenna.

\begin{center}
\begin{figure}[t]
\includegraphics[width=8cm]{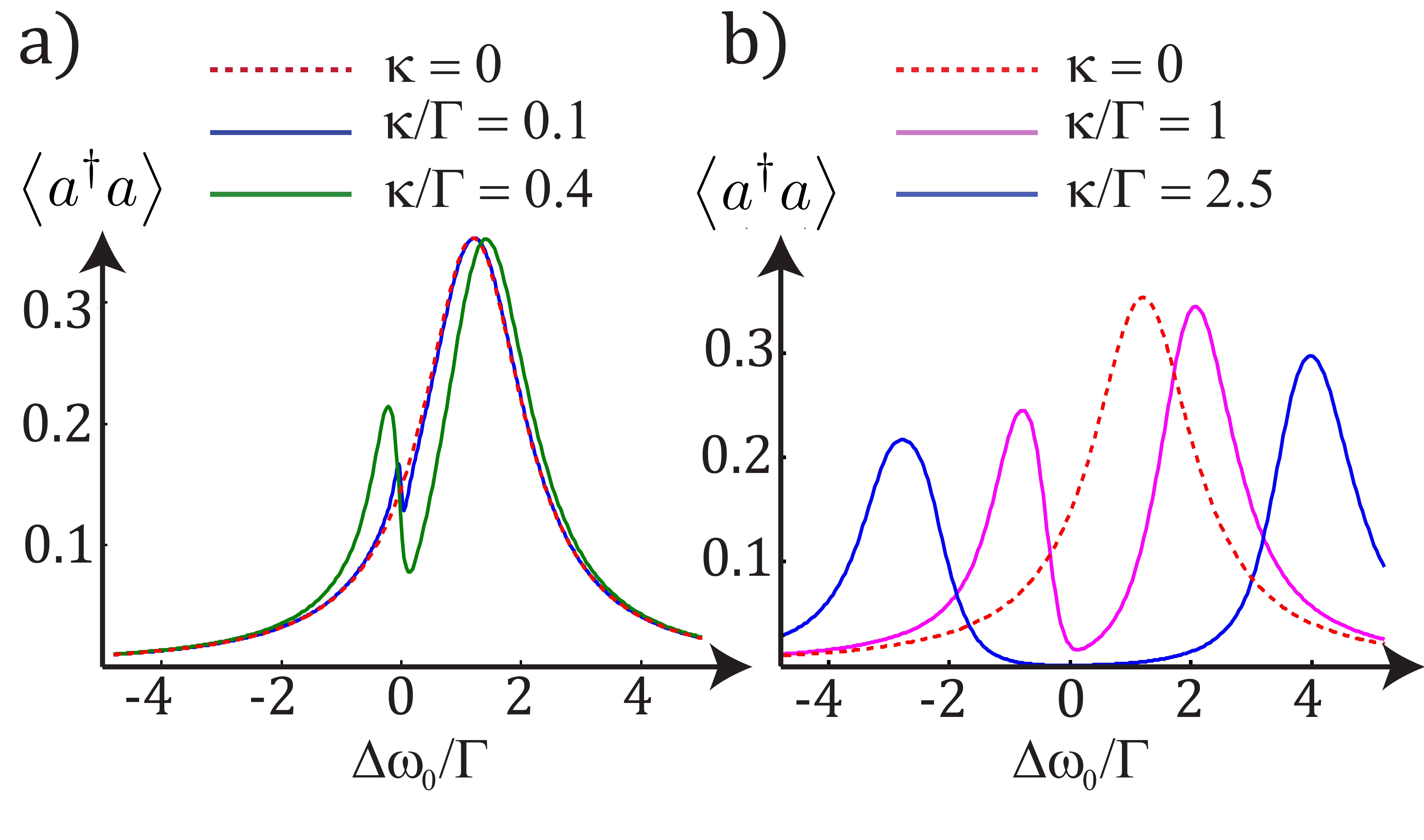} \begin{caption}{Impact of atoms
on the extinction spectra of the hybrid system for different coupling
strengths $\kappa$ in the weak (a) and strong coupling regime (b).
The steady-state mean number of photons is plotted vs. the driving-field
normalized detuning from the atomic transition frequency $\Delta\omega_{0}/\Gamma$.
For the sake of comparison the spectrum of the bare nanoantenna (no
coupling, $\kappa=0$) has been added. A Fano-like behavior can be
observed for increased but still weak coupling. The results have been
obtained for a weak driving field (Rabi frequency $\Omega=0.6\Gamma$).
Furthermore, a rather small detuning between nanoantenna resonance
and atomic transition frequency ($\omega_{\mathrm{na}}-\omega_{0}=1.2\Gamma$)
has been assumed. The blue line in panel (b) corresponds to the antenna
design of Sec. \ref{sec:design}. \label{fig:spectra_kappa}} \end{caption}
\end{figure}

\par\end{center}

Generally, the spectrum can be understood in terms of hybridization
caused by the interaction of all sub-systems \cite{Meystre1999}.
The eigenstates and eigenenergies can be derived on the basis of the
Jaynes-Cummings model which is outlined in App.~\ref{app:eigenstates}
in detail. The eigenenergies are plotted in Fig.~\ref{fig:diagrams}
for the cases of a single and two atoms coupled to the nanoantenna.
In the resonant case ($\omega_{\mathrm{na}}=\omega_{0}$) the splitting
between the first pair of excited eigenstates is equal to $\sqrt{N_{\mathrm{tls}}}\kappa$.
This simple example proves that the spectrum of the hybrid system,
i.e. the position of the Rabi peaks, crucially depends on the number
of atoms. Furthermore, the effective increase of the coupling constant
by the factor of $\sqrt{N_{\mathrm{tls}}}$ may help to overcome losses
and fulfill the strong coupling condition (\ref{eq:condition_strong_coupling}).

The analysis of the diagrams in Fig. \ref{fig:diagrams} suggests
that, in principle, one might expect an even stronger manifestation
of the number of atoms in the spectra for slightly more intense driving
fields, where highly-excited states ($n\geq2$) become occupied. For
this purpose, however, a coupling even stronger than that obtained
with the design of Sec. \ref{sec:design} would be preferable. Otherwise,
due to significant nanoantenna losses the required sensitivity to
trace the contribution of highly excited states to the spectra cannot
be reached.

\begin{center}
\begin{figure}[H]
\includegraphics[width=8cm]{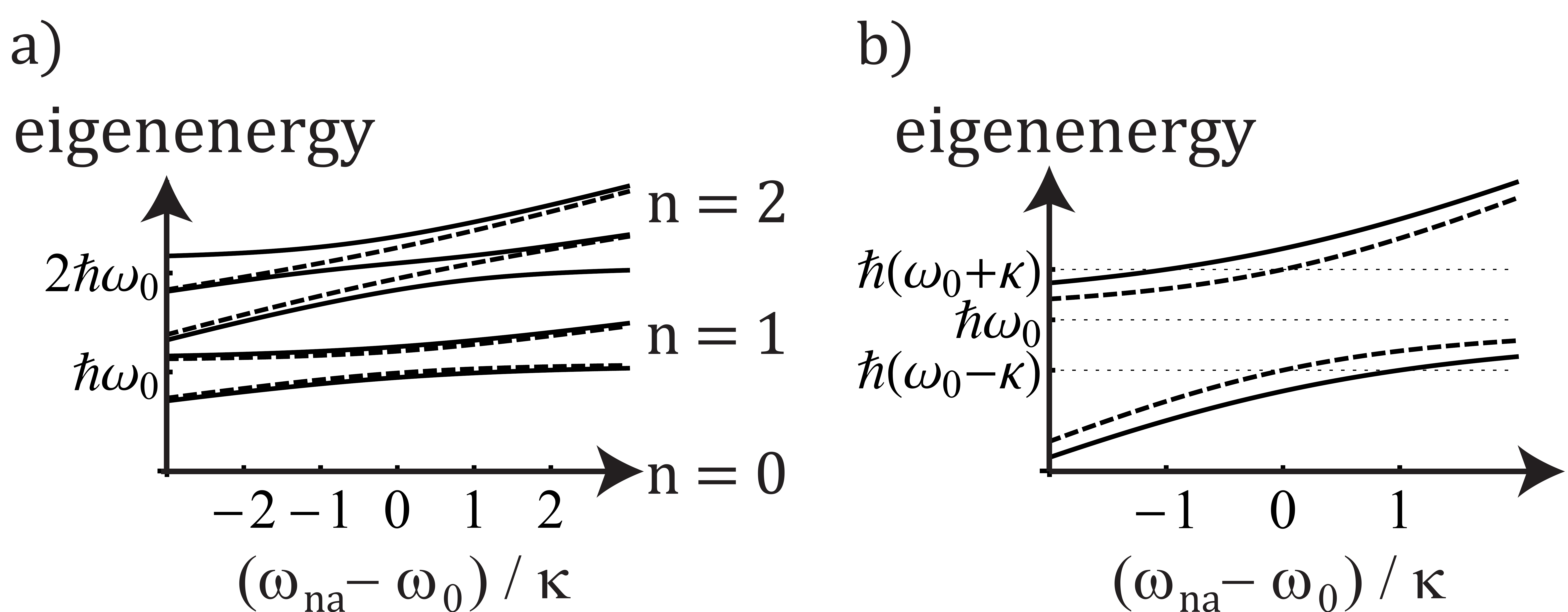} \begin{caption}{(a) Eigenenergies
of the hybrid system as function of the detuning between nanoantenna
resonance and atomic transition frequency for the case of $N_{\mathrm{tls}}=1$
(dashed lines) and $N_{\mathrm{tls}}=2$ (solid lines). Only energies
of states corresponding to the total number of excitations $n\leq2$
are shown. (b) Zoomed view for $n=1$. Energy splitting of the first
pair of excited states compared for the cases $N_{\mathrm{tls}}=1$
and $N_{\mathrm{tls}}=2$. In the latter case it is larger by a factor
of $\sqrt{2}$ for $\omega_{\mathrm{na}}=\omega_{0}$. \label{fig:diagrams}}
\end{caption}
\end{figure}

\par\end{center}

The interesting question arises whether the influence of the atoms
on the extinction spectra remains so significant for stronger driving
fields, i.e. beyond the quantum regime. To investigate such transition
we performed calculations for increasing driving intensities for the
strong-coupling design proposed in Section \ref{sec:design}. The
results are displayed in Fig.~\ref{fig:spectra_omega}(a). For weak
driving fields (blue line) a splitting can be observed with Rabi peaks
at $\Delta\omega_{0}=\pm\sqrt{2}\kappa$ as well as an onset of transparency
at $\Delta\omega_{0}=0$, as measured in cavity QED systems, see Refs.~\cite{Hood1998,Miller2005}.
For stronger driving fields the corresponding extinction peaks are
shifted towards the center. Finally, in the limit of a strong driving
field ($\Omega>\Gamma$), the contribution from the atomic system
becomes negligible and the bare nanoantenna spectrum is recovered.

To understand this behavior, we analyze the occupation probabilities
of the hybrid system's eigenstates. For weak driving fields only the
first pair of excited eigenstates $|\psi_{1,\pm}\rangle$ with energies
equal to $\hbar\omega_{0}\pm\hbar\sqrt{2}\kappa$ is populated. For
this case the occupation probabilities of the states $|\psi_{1,+}\rangle$
and $|\psi_{2,+}\rangle$ are plotted in Fig.~\ref{fig:spectra_omega}(b).
It can be seen that for the latter state it is indeed negligible.
The occupation probabilities of states $|\psi_{1,-}\rangle$ and $|\psi_{2,-}\rangle$
(not shown) are peaks symmetric with respect to $|\psi_{1,+}\rangle$
and $|\psi_{2,+}\rangle$, i.e. they are centered at $\omega=\omega_{0}+\sqrt{2}\kappa$.
For an increased driving field, the probability of exciting
higher-energy states becomes significant {[}see green lines in Fig.~\ref{fig:spectra_omega}(b){]}.
This is the reason for the shift of the Rabi peaks \cite{Meystre1999}
towards the center. If the driving field rises, the energy difference
between subsequent occupied eigenstates converges towards $\omega_{\mathrm{na}}$,
which explains why in the strong driving field limit ($\Omega>\Gamma$)
the result corresponds to the bare nanoantenna case. This result can
also be derived from the steady-state solution of the Heisenberg equations
of motion where one finds that $\langle a^{\dagger}a\rangle\approx\Omega^{2}/\Gamma^{2}\gg N_{\mathrm{tls}}$
holds. This means that indeed for stronger driving fields the classical
behavior of the nanoantenna is re-established, irrespective of the
presence of the atoms.

\begin{center}
\begin{figure}[H]
\includegraphics[width=8cm]{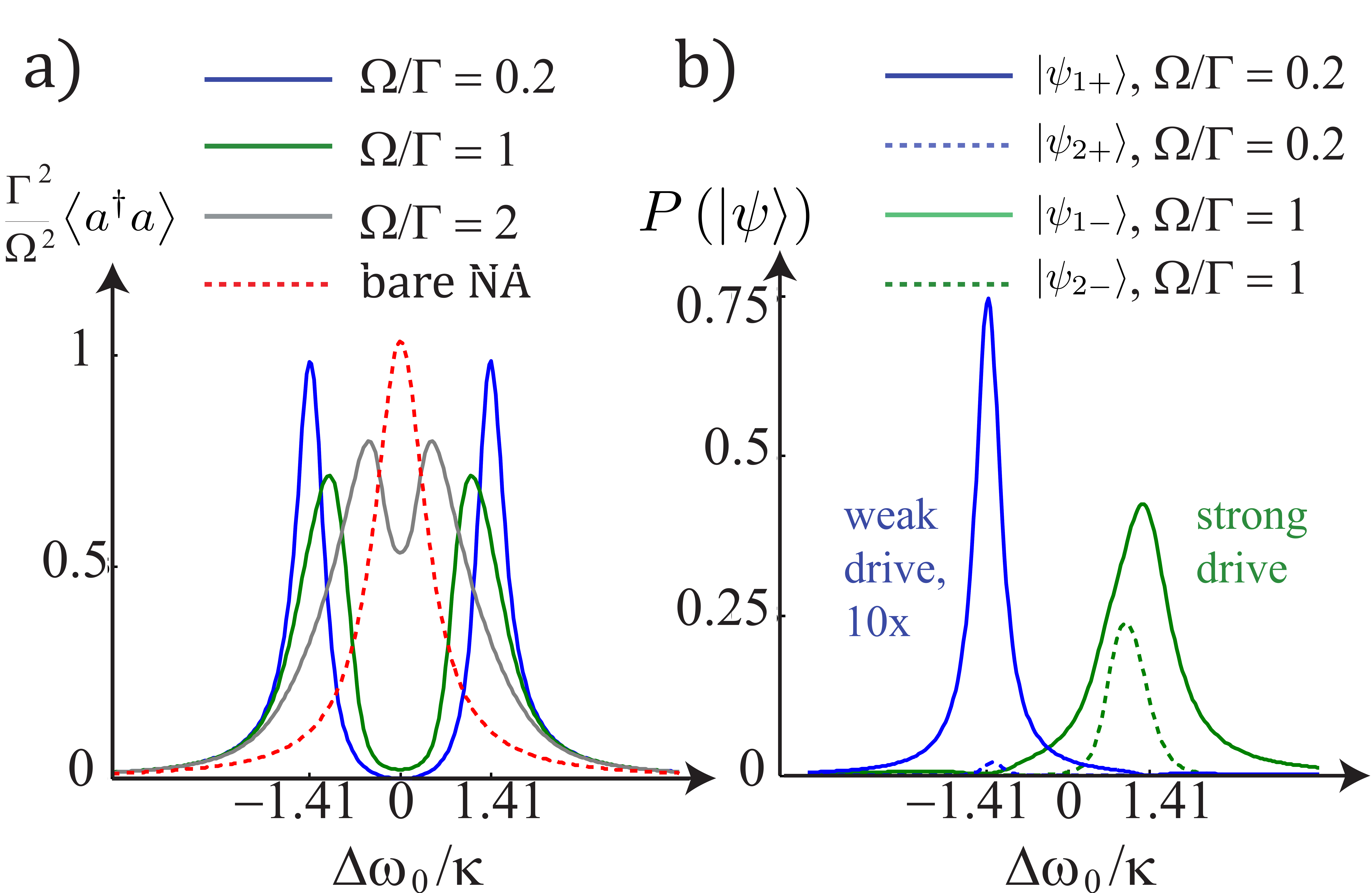} \begin{caption}{(a) Impact of
the driving field intensity on the extinction spectra, normalized
by a dimensionless parameter $\Omega^{2}/\Gamma^{2}$, for the strongly
coupled nanoantenna proposed in Sec. \ref{sec:design}. (b) Probability
of occupation of the first ($n=1$, solid lines) and second ($n=2$,
dashed lines) pair of excited states $|\psi_{n,\pm}\rangle$ for a
driving field with Rabi frequencies $\Omega=0.2\Gamma$ (blue lines,
results multiplied by a factor of $10$) and $\Omega=\Gamma$ (green
lines). \label{fig:spectra_omega}} \end{caption}
\end{figure}

\par\end{center}

\section{Conclusions\label{sec:concl}}

In this paper, we have investigated the coupling of one and two atoms
approximated by two-level systems to optical nanoantennas. It was
outlined that a full quantum approach is required to understand the
dynamics of the hybrid system. Only for extremely weak driving fields
a semiclassical formulation of the nanoantenna dynamics may be used.

At the design stage for a nanoantenna a trade-off must be taken into
account between the field enhancement by the nanoantenna, directly
responsible for the strength of the coupling, and the nanoantenna's
efficiency. Small size, which results in high absorption losses, is
essential for achieving the strong coupling regime. Then, the spectra
of the hybrid system are hugely influenced by the presence of the
atoms provided that the driving field is considerably weak, i.e. in
the quantum regime.

Furthermore we have shown that a strong coupling of several atoms
to a nanoantenna does change the spectrum significantly. Such features
will enable experimentalists to identify situations of multiple coupled
atoms in the strong coupling regime.

\section*{Acknowledgement}

This work was partially supported by the German Federal Ministry of
Education and Research (PhoNa) and by the Thuringian State Government
(MeMa).

\appendix

\section{The dynamics of the classical field\label{app:class_field_dynamics}}

In this section we will outline the semiclassical treatment of the
nanoantenna used in this paper. Especially, we will derive the equations
of motion for the respective electric field in dependence on the state
of the atoms.

If a plasmonic structure like the discussed nanoantenna is much smaller
than the wavelength, it can be treated as a classical harmonic oscillator
as well-known in the metamaterials community, see e.g. Ref.~\cite{Petschulat2008}.
The scattered field $\tilde{\mathbf{E}}_{\mathrm{exc}}\left(\mathbf{r},t\right)$
of such an oscillator may be separated into temporally and spatially
varying contributions:
\begin{eqnarray}
\tilde{\mathbf{E}}_{\mathrm{exc}}\left(\mathbf{r},t\right) & = & \tilde{\alpha}\left(t\right)\mathbf{E}_{\mathrm{exc}}\left(\mathbf{r}\right)+\mathrm{c.c.}\ ,
\end{eqnarray}
where the subscript \textit{\emph{'exc'}} is used to label the excited
field of the nanoantenna. This notation renders it unnecessary to
distinguish between the scattered field of the nanoantenna and the
field inside its metallic body. The spatial part $\mathbf{E}_{\mathrm{exc}}\left(\mathbf{r}\right)$
can be found by computer simulations or analytical considerations.
By definition we assume that this spatial contribution, which may
also be denoted as mode profile, does not change with time for the
excitations we discuss in the following. Then, the whole nanoantenna
dynamics is exclusively described by the evolution of the temporal
part $\tilde{\alpha}\left(t\right)$ for which the equations of motion
will be derived and solved in the following.

Under the assumption of an oscillator-like evolution, the positive
frequency part $\tilde{\alpha}\left(t\right)$ evolves according to
the equation
\begin{eqnarray}
\ddot{\tilde{\alpha}}\left(t\right)+\Gamma\dot{\tilde{\alpha}}\left(t\right)+\omega_{\mathrm{na}}^{2}\tilde{\alpha}\left(t\right) & = & F\left(t\right)e^{-i\omega_{\mathrm{dr}}t},
\end{eqnarray}
where $F\left(t\right)e^{-i\omega_{\mathrm{dr}}t}$ is the driving
field at the nanoantenna's site. In our case it is proportional to
the driving laser field and the assumed dipolar fields of the two
atoms.

We assume the driving term to oscillate at the mean frequency $\omega_{\mathrm{dr}}$.
Then, its envelope $F\left(t\right)$ varies much slower. Consequentially,
we can also calculate the solutions to the equations of motion for
the slowly-varying part of the nanoantennas oscillation defined via
$\tilde{\alpha}\left(t\right)=\alpha\left(t\right)e^{-i\omega_{\mathrm{dr}}t}$.
Then using the standard slowly-varying envelope approximation ( $\left|\dot{\alpha}\right|\ll\left|\alpha\omega_{\mathrm{dr}}\right|$)
and accounting for $\Gamma\ll\omega_{\mathrm{dr}}$ one arrives at
the equations of motion for the electric field of the nanoantenna
as
\begin{equation}
\dot{\alpha}\left(t\right)=-\Gamma\alpha\left(t\right)+\frac{i}{2\omega_{\mathrm{dr}}}\left[\left(\omega_{\mathrm{dr}}^{2}-\omega_{\mathrm{na}}^{2}\right)\alpha(t)+F\left(t\right)\right].
\end{equation}
For near-resonance driving fields ($\omega_{\mathrm{na}}^{2}-\omega_{\mathrm{dr}}^{2}\approx2\omega_{\mathrm{dr}}\Delta\omega_{\mathrm{na}}$),
the above equation further reduces to:
\begin{equation}
\dot{\alpha}\left(t\right)=-\left(\Gamma+i\Delta\omega_{\mathrm{na}}\right)\alpha\left(t\right)+iF\left(t\right)/2\omega_{\mathrm{dr}}\ ,\label{eq:alpha_with_F}
\end{equation}
We now compare equation (\ref{eq:alpha_with_F}) with the Heisenberg
operator equation in the fully quantum approach, \mbox{$\dot{a}\left(t\right)=-\left(\Gamma+i\Delta\omega_{\mathrm{na}}\right)a\left(t\right)+i\left[\kappa\sum_{j}\sigma_{-}^{(j)}\left(t\right)+\Omega\right]+f(t)$},
where $f(t)$ stands for the Langevin noise operator such that $\langle f(t)\rangle=0$
\cite{Meystre1999}, which origins from coupling of the field of the
nanoantenna with its electromagnetic environment and with phonons.
With such a direct comparison, we can identify the driving term
\begin{eqnarray}
F\left(t\right) & = & 2\omega_{\mathrm{dr}}\left[\kappa\sum_{j}{\rho_{eg}^{\mathrm{sc}}}^{(j)}(t)+\Omega\right]\ ,\ \mathrm{so}
\end{eqnarray}
and get eventually
\begin{equation}
\dot{\alpha}\left(t\right)=-\left(\Gamma+i\Delta\omega_{\mathrm{na}}\right)\alpha\left(t\right)+i\left[\kappa\sum_{j}{\rho_{eg}^{\mathrm{sc}}}^{(j)}(t)+\Omega\right]
\end{equation}
as the evolution equation for the electric field of the nanoantenna
in the semiclassical approximation. Our result corresponds to the
equation of motion of the expectation value of quantum operators if
the nanoantenna's field is approximated by a coherent state $|\alpha\left(t\right)\rangle$.

\section{Calculation of parameters\label{app:coupling_constants}}

In Appendix A we have derived the equations of motion for the electric
field of a nanoantenna coupled to an unspecified number of noninteracting
atoms. Now we will specify how to obtain the relevant system parameters
from electromagnetic simulations. The key for determining the coupling
constants and loss rates lies in the determination of the nanoantenna
field for a single excitation as we will see shortly.

\subsection*{Coupling constant between \protect \protect \\
 nanoantenna and atoms: $\kappa$ \label{sub:kappa}}

In electric dipole approximation the parameter $\kappa$ describing
the coupling strength between the $j^{\mathrm{th}}$ atom and the
nanoantenna is given by
\begin{equation}
\kappa_{j}=\mathbf{E}_{\mathrm{exc}}^{\mathrm{se}}\left(\mathbf{r}_{j}\right)\cdot\mathbf{d}_{ge}^{(j)}/\hbar\ ,
\end{equation}
where $\mathbf{d}_{ge}^{(j)}$ stands for the transition dipole moment
in the $j^{\mathrm{th}}$ atom positioned at $\mathbf{r}_{j}$. In the considered case the atoms are
assumed to be identical, so the index $j$ can be dropped. Naturally,
here the field $\mathbf{E}_{\mathrm{exc}}^{\mathrm{se}}$ corresponds
to a single excitation of the nanoantenna subject to a plane-wave
excitation. Because of the atomic system's symmetric positioning and
the mirror-symmetry of the nanoantenna, $\mathbf{E}_{\mathrm{exc}}^{\mathrm{se}}\left(\mathbf{r}_{1}\right)=\mathbf{E}_{\mathrm{exc}}^{\mathrm{se}}\left(\mathbf{r}_{2}\right)$
 the excitation of the antisymmetric
state of the atomic system for the lossless case is prohibited, see
App.~\ref{app:eigenstates}. It is evident that $\kappa$ depends
on the dipole moment of both atoms, the electric field of the nanoantenna's
mode and the very location of the atoms. Obviously, this scheme of
calculating the field of the nanoantenna corresponds to a dipolar
mode approximation since the plane wave mainly couples to the dipolar
mode of the nanoantenna. Higher order contributions are neglected\cite{Yan2008},
which has two consequences: the coupling strengths $\kappa_{j}$ as
well as the nonradiative loss rate of the nanoantenna, $\Gamma_{\mathrm{nr}}$,
are underestimated. We have chosen to restrict our investigations
to this simplified scheme as the inclusion of higher order nanoantenna
modes significantly complicates the analysis of the system dynamics
and also prevents us from finding accessible analytical results.

To find the correct scaling for $\kappa$, the electric field has
to be calculated at the positions of the atoms for an excitation of
the nanoantenna by a single photon with energy $\hbar\omega_{\mathrm{na}}$.
Thus, for the computation at the nanoantenna's resonance, first the
field energy for the corresponding excited electromagnetic mode $\mathbf{E}_{\mathrm{exc}}\left(\mathbf{r}\right)$
has to be determined using the well-known energy-density integration
for dispersive media \cite{Landau}
\begin{eqnarray}
W & = & \frac{1}{2}\int\left.\frac{\partial}{\partial\omega}\left[\omega\,\Re\varepsilon\left(\omega\right)\right]\right|_{\omega=\omega_{\mathrm{na}}}\left|E_{\mathrm{exc}}\left(\mathbf{r}\right)\right|^{2}dV\\
 &  & +\frac{1}{2}\int\mu_{0}\left|H_{\mathrm{exc}}\left(\mathbf{r}\right)\right|^{2}dV\ .\nonumber
\end{eqnarray}
Then, the electric field of a single-photon excitation is given by
\begin{eqnarray}
\mathbf{E}_{\mathrm{exc}}^{\mathrm{se}}\left(\mathbf{r}\right) & = & \sqrt{\hbar\omega_{\mathrm{na}}/W}\,\mathbf{E}_{\mathrm{exc}}\left(\mathbf{r}\right)\ ,
\end{eqnarray}
since the energy of the nanoantenna's mode corresponds to $N=W/\hbar\omega_{\mathrm{na}}$
photons.

\subsection*{Coupling constant between nanoantenna \protect \\
and driving field: Rabi frequency $\Omega$}

The Rabi frequency $\Omega$ describes the coupling of the nanoantenna
and the driving field. It can be evaluated in the dipole approximation
as $\Omega=\mathbf{d}_{\mathrm{na}}^{\mathrm{se}}\cdot\mathbf{E}_{\mathrm{dr}}/\hbar$,
where $\mathbf{d}_{\mathrm{na}}^{\mathrm{se}}$ is the dipole moment
of the nanoantenna corresponding to a single excitation for which
the electric field is known from the previous subsection. The dipole
moment $\mathbf{d}_{\mathrm{na}}^{\mathrm{se}}$ can then just be
calculated from a multipole expansion of $\mathbf{E}_{\mathrm{exc}}^{\mathrm{se}}\left(\mathbf{r}\right)$.
For our calculations, the main contribution to the far-field of the
nanoantenna was indeed that from the dipole moment which justifies
the calculation of $\Omega$ in the dipole approximation.

\subsection*{Radiative and nonraditive losses \protect \protect \\
 of the nanoantenna: $\Gamma$}

The losses of the nanoantenna can be divided into radiative and non-radiative
losses, $\Gamma=\Gamma_{\mathrm{r}}+\Gamma_{\mathrm{nr}}$. Both quantities
are related to integrations of the nanoantenna mode for a single excitation
$\mathbf{E}_{\mathrm{exc}}^{\mathrm{se}}\left(\mathbf{r}\right)$:
The radiative loss $\Gamma_{\mathrm{r}}$ can be determined by integrating
the time-averaged Poynting vector over a closed surface embedding
the nanoantenna, $\Gamma_{\mathrm{r}}=\int\left\langle \mathbf{S}_{\mathrm{exc}}^{\mathrm{se}}\left(\mathbf{r},t\right)\right\rangle d\mathbf{A}$.
The non-radiative part is given by a volume integral over the nanoantenna
using Ohm's law, $\Gamma_{\mathrm{nr}}=\int\sigma\left\langle \mathbf{E}_{\mathrm{exc}}^{\mathrm{se}}\left(\mathbf{r},t\right)\right\rangle ^{2}dV$,
where $\sigma$ is the electric conductivity of the metal.

\section{Eigenstates of the free Hamiltonian\label{app:eigenstates}}

Now we are going to analyze the Hamiltonian of the nanoantenna coupled
to two two-level systems in the fully quantum case. As one might expect,
we will arrive at a simple generalization of the Jaynes-Cummings model
and give the energy spectrum of the hybrid system's eigenstates.

It is often advantageous to analyze problems of coupled two-level
systems in the so-called Dicke basis that relates ground and excited
states, $|g\rangle$ and $|e\rangle$, of each two-level system to
a combined eigenbasis \cite{Dicke1954}: $\left\{ |D\rangle\equiv|e\rangle\otimes|e\rangle\right.$,
$|S\rangle\equiv\frac{1}{\sqrt{2}}\left(|e\rangle\otimes|g\rangle+|g\rangle\otimes|e\rangle\right)$,
$|A\rangle\equiv\frac{1}{\sqrt{2}}\left(-|e\rangle\otimes|g\rangle+|g\rangle\otimes|e\rangle\right)$,
$\left.|G\rangle\equiv|g\rangle\otimes|g\rangle\right\} $, with the
doubly excited state $|D\rangle$, the symmetric and antisymmetric
states $|S\rangle$ and $|A\rangle$ with a single excitation, and
the ground state $|G\rangle$. In the Dicke basis, the Hamiltonian
reads as
\begin{eqnarray}
H & = & \frac{1}{2}\hbar\omega_{0}\left(2|D\rangle\langle D|+|S\rangle\langle S|+|A\rangle\langle A|\right)+\hbar\omega_{\mathrm{na}}a^{\dagger}a\nonumber \\
 &  & -\hbar\sqrt{2}\kappa\left(\Sigma_{+}a+a^{\dagger}\Sigma_{-}\right)\ ,
\end{eqnarray}
where
\begin{eqnarray*}
\Sigma_{+} & = & \frac{1}{\sqrt{2}}\left(\sigma_{+}^{(1)}+\sigma_{+}^{(2)}\right)=|D\rangle\langle S|+|S\rangle\langle G|\ \mathrm{and}\\
\Sigma_{-} & = & \Sigma_{+}^{\dagger},
\end{eqnarray*}
are the creation and annihilation operators of an excitation in the
atomic subsystem. Note that the antisymmetric state is decoupled in
the isolated system and can be populated only by decay mechanisms
or by an asymmetric drive. From now on it suffices to consider only
the effective three-level system, whose state belongs to the Hilbert
space spanned by $\{|G\rangle,|S\rangle,|D\rangle\}$.

We will give the explicit form of the eigenstates of the Hamiltonian
and the corresponding eigenenergies in the case of resonance between
the atomic and plasmonic systems, i.e. when $\omega_{\mathrm{na}}=\omega_{0}$.
The states of the hybrid system can be expressed in the Dicke basis
for the atomic subsystem, and in the Fock basis $\{|k\rangle\}_{k=0}^{\infty}$
for the nanoantenna, with $k$ denoting the number of photons in the
system.

Each state can be characterized by the total number of excitations
$n$ it corresponds to. For instance, $n=0$ stands for the total
ground state of the system $|\psi_{0}\rangle=|G,0\rangle$ of energy
$E_{0}=0$. For a single excitation $n=1$ we have two eigenstates
and eigenvalues:
\begin{eqnarray}
|\psi_{1,\pm}\rangle & = & \pm|S,0\rangle+|G,1\rangle,\\
E_{1,\pm} & = & \hbar\omega_{0}\mp\sqrt{2}\hbar\kappa.
\end{eqnarray}
For the numbers of excitations $n\geq2$ there are three eigenstates
for a given $n$:
\begin{eqnarray}
|\psi_{n,\pm}\rangle & = & \sqrt{n-1}|D,n-2\rangle\\
 &  & \pm\sqrt{2n-1}|S,n-1\rangle+\sqrt{n}|G,n\rangle,\nonumber \\
E_{n,\pm} & = & n\hbar\omega_{0}\mp\sqrt{2\left(2n-1\right)}\hbar\kappa,\\
|\psi_{n,0}\rangle & = & \sqrt{n}|D,n-2\rangle-\sqrt{n-1}|G,n\rangle,\\
E_{n,0} & = & n\hbar\omega_{0}.
\end{eqnarray}
For visibility, the states are not normalized. Note that the eigenstates
of the total system cannot be written as a product of states of atomic
and nanoantenna subsystems. This is a clear sign of strong interaction
of the subsystems that leads to their entanglement. The latter is
a sole quantum feature which cannot be accounted for with the semiclassical
formalism. Our condition for the validity of the semiclassical approach
can now be seen from a different perspective: the semiclassical description
can be applied only if entanglement is present in the system with
negligible probability.

The eigenstates attain a more complicated form, if the two-level systems
are not in resonance with the nanoantenna ($\omega_{\mathrm{na}}\neq\omega_{0}$).
Then, the energy diagram depends strongly on the detuning, see again
Fig.~\ref{fig:diagrams}. In the strongly off-resonant limit, the
interaction becomes negligible and the atomic and the nanoantenna
subsystems behave independently. Consequentially, the eigenenergies
converge towards the unperturbed values.

For large numbers of excitations, the eigenstates become approximately
separable:
\begin{eqnarray}
|\psi_{n,\pm}\rangle & \approx & \left(|D\rangle\pm\sqrt{2}|S\rangle+|G\rangle\right)\otimes|n\rangle,\\
|\psi_{n,0}\rangle & \approx & \left(|D\rangle-|G\rangle\right)\otimes|n\rangle,
\end{eqnarray}
with the interaction energies: \mbox{$\Delta E_{n,\pm}=\mp2\hbar\kappa\sqrt{n}$},
\mbox{$\Delta E_{n,0}=0$}. This means, that in the limit of large
field intensities, even though the field has strong influence on the
atoms, the atoms approximately do not affect the field and the semiclassical
approach can be applied again.

\bibliography{strong_coupling_september}

\begin{thebibliography}{54}
\expandafter\ifx\csname natexlab\endcsname\relax\def\natexlab#1{#1}\fi
\expandafter\ifx\csname bibnamefont\endcsname\relax
  \def\bibnamefont#1{#1}\fi
\expandafter\ifx\csname bibfnamefont\endcsname\relax
  \def\bibfnamefont#1{#1}\fi
\expandafter\ifx\csname citenamefont\endcsname\relax
  \def\citenamefont#1{#1}\fi
\expandafter\ifx\csname url\endcsname\relax
  \def\url#1{\texttt{#1}}\fi
\expandafter\ifx\csname urlprefix\endcsname\relax\def\urlprefix{URL }\fi
\providecommand{\bibinfo}[2]{#2}
\providecommand{\eprint}[2][]{\url{#2}}

\bibitem[{\citenamefont{Anger et~al.}(2006)\citenamefont{Anger, Bharadwaj, and
  Novotny}}]{Anger2006}
\bibinfo{author}{\bibfnamefont{P.}~\bibnamefont{Anger}},
  \bibinfo{author}{\bibfnamefont{P.}~\bibnamefont{Bharadwaj}},
  \bibnamefont{and} \bibinfo{author}{\bibfnamefont{L.}~\bibnamefont{Novotny}},
  \bibinfo{journal}{Phys. Rev. Lett.} \textbf{\bibinfo{volume}{96}},
  \bibinfo{pages}{113002} (\bibinfo{year}{2006}).

\bibitem[{\citenamefont{Pfeiffer et~al.}(2010)\citenamefont{Pfeiffer, Lindfors,
  Wolpert, Atkinson, Benyoucef, Rastelli, Schmidt, Giessen, and
  Lippitz}}]{Giessen2010}
\bibinfo{author}{\bibfnamefont{M.}~\bibnamefont{Pfeiffer}},
  \bibinfo{author}{\bibfnamefont{K.}~\bibnamefont{Lindfors}},
  \bibinfo{author}{\bibfnamefont{C.}~\bibnamefont{Wolpert}},
  \bibinfo{author}{\bibfnamefont{P.}~\bibnamefont{Atkinson}},
  \bibinfo{author}{\bibfnamefont{M.}~\bibnamefont{Benyoucef}},
  \bibinfo{author}{\bibfnamefont{A.}~\bibnamefont{Rastelli}},
  \bibinfo{author}{\bibfnamefont{O.~G.} \bibnamefont{Schmidt}},
  \bibinfo{author}{\bibfnamefont{H.}~\bibnamefont{Giessen}}, \bibnamefont{and}
  \bibinfo{author}{\bibfnamefont{M.}~\bibnamefont{Lippitz}},
  \bibinfo{journal}{Nano Letters} \textbf{\bibinfo{volume}{10}},
  \bibinfo{pages}{4555} (\bibinfo{year}{2010}).

\bibitem[{\citenamefont{Curto et~al.}(2010)\citenamefont{Curto, Volpe,
  Taminiau, Kreuzer, Quidant, and van Hulst}}]{vanHulst2010}
\bibinfo{author}{\bibfnamefont{A.~G.} \bibnamefont{Curto}},
  \bibinfo{author}{\bibfnamefont{G.}~\bibnamefont{Volpe}},
  \bibinfo{author}{\bibfnamefont{T.~H.} \bibnamefont{Taminiau}},
  \bibinfo{author}{\bibfnamefont{M.~P.} \bibnamefont{Kreuzer}},
  \bibinfo{author}{\bibfnamefont{R.}~\bibnamefont{Quidant}}, \bibnamefont{and}
  \bibinfo{author}{\bibfnamefont{N.~F.} \bibnamefont{van Hulst}},
  \bibinfo{journal}{Science} \textbf{\bibinfo{volume}{329}},
  \bibinfo{pages}{930} (\bibinfo{year}{2010}).

\bibitem[{\citenamefont{Zhu et~al.}(2012)\citenamefont{Zhu, Xie, Shi, Liu,
  Mortensen, Xiao, Zi, and Choy}}]{Zhu2012}
\bibinfo{author}{\bibfnamefont{X.}~\bibnamefont{Zhu}},
  \bibinfo{author}{\bibfnamefont{F.}~\bibnamefont{Xie}},
  \bibinfo{author}{\bibfnamefont{L.}~\bibnamefont{Shi}},
  \bibinfo{author}{\bibfnamefont{X.}~\bibnamefont{Liu}},
  \bibinfo{author}{\bibfnamefont{N.~A.} \bibnamefont{Mortensen}},
  \bibinfo{author}{\bibfnamefont{S.}~\bibnamefont{Xiao}},
  \bibinfo{author}{\bibfnamefont{J.}~\bibnamefont{Zi}}, \bibnamefont{and}
  \bibinfo{author}{\bibfnamefont{W.}~\bibnamefont{Choy}},
  \bibinfo{journal}{Opt. Lett.} \textbf{\bibinfo{volume}{37}},
  \bibinfo{pages}{2037} (\bibinfo{year}{2012}).

\bibitem[{\citenamefont{Filter et~al.}(2013)\citenamefont{Filter, Farhat,
  Steglich, Alaee, Rockstuhl, and Lederer}}]{Filter2013Gr}
\bibinfo{author}{\bibfnamefont{R.}~\bibnamefont{Filter}},
  \bibinfo{author}{\bibfnamefont{M.}~\bibnamefont{Farhat}},
  \bibinfo{author}{\bibfnamefont{M.}~\bibnamefont{Steglich}},
  \bibinfo{author}{\bibfnamefont{R.}~\bibnamefont{Alaee}},
  \bibinfo{author}{\bibfnamefont{C.}~\bibnamefont{Rockstuhl}},
  \bibnamefont{and} \bibinfo{author}{\bibfnamefont{F.}~\bibnamefont{Lederer}},
  \bibinfo{journal}{Optics Express} \textbf{\bibinfo{volume}{21}},
  \bibinfo{pages}{3737} (\bibinfo{year}{2013}).

\bibitem[{\citenamefont{Mohtashami and Koenderink}(2013)}]{Mohtashami2013}
\bibinfo{author}{\bibfnamefont{A.}~\bibnamefont{Mohtashami}} \bibnamefont{and}
  \bibinfo{author}{\bibfnamefont{A.~F.} \bibnamefont{Koenderink}},
  \bibinfo{journal}{New J. Phys.} \textbf{\bibinfo{volume}{15}},
  \bibinfo{pages}{043017} (\bibinfo{year}{2013}).

\bibitem[{\citenamefont{Novotny and van Hulst}(2011)}]{novotny2011antennas}
\bibinfo{author}{\bibfnamefont{L.}~\bibnamefont{Novotny}} \bibnamefont{and}
  \bibinfo{author}{\bibfnamefont{N.}~\bibnamefont{van Hulst}},
  \bibinfo{journal}{Nat. Phot.} \textbf{\bibinfo{volume}{5}},
  \bibinfo{pages}{83} (\bibinfo{year}{2011}).

\bibitem[{\citenamefont{Rogobete et~al.}(2007)\citenamefont{Rogobete, Kaminski,
  Agio, and Sandoghdar}}]{Rogobete2007}
\bibinfo{author}{\bibfnamefont{L.}~\bibnamefont{Rogobete}},
  \bibinfo{author}{\bibfnamefont{F.}~\bibnamefont{Kaminski}},
  \bibinfo{author}{\bibfnamefont{M.}~\bibnamefont{Agio}}, \bibnamefont{and}
  \bibinfo{author}{\bibfnamefont{V.}~\bibnamefont{Sandoghdar}},
  \bibinfo{journal}{Opt. Lett.} \textbf{\bibinfo{volume}{32}},
  \bibinfo{pages}{1623} (\bibinfo{year}{2007}).

\bibitem[{\citenamefont{Kern et~al.}(2012)\citenamefont{Kern, Grossmann,
  Tarakina, H\"{a}ckel, Emmerling, Kamp, Huang, Biagioni, Prangsma, and
  Hecht}}]{Hecht2012}
\bibinfo{author}{\bibfnamefont{J.}~\bibnamefont{Kern}},
  \bibinfo{author}{\bibfnamefont{S.}~\bibnamefont{Grossmann}},
  \bibinfo{author}{\bibfnamefont{N.~V.} \bibnamefont{Tarakina}},
  \bibinfo{author}{\bibfnamefont{T.}~\bibnamefont{H\"{a}ckel}},
  \bibinfo{author}{\bibfnamefont{M.}~\bibnamefont{Emmerling}},
  \bibinfo{author}{\bibfnamefont{M.}~\bibnamefont{Kamp}},
  \bibinfo{author}{\bibfnamefont{J.-S.} \bibnamefont{Huang}},
  \bibinfo{author}{\bibfnamefont{P.}~\bibnamefont{Biagioni}},
  \bibinfo{author}{\bibfnamefont{J.~C.} \bibnamefont{Prangsma}},
  \bibnamefont{and} \bibinfo{author}{\bibfnamefont{B.}~\bibnamefont{Hecht}},
  \bibinfo{journal}{Nano Lett.} \textbf{\bibinfo{volume}{12}},
  \bibinfo{pages}{5504} (\bibinfo{year}{2012}).

\bibitem[{\citenamefont{Kern and Martin}(2012)}]{Kern2012}
\bibinfo{author}{\bibfnamefont{A.~M.} \bibnamefont{Kern}} \bibnamefont{and}
  \bibinfo{author}{\bibfnamefont{O.~J.~F.} \bibnamefont{Martin}},
  \bibinfo{journal}{Phys. Rev. A} \textbf{\bibinfo{volume}{85}},
  \bibinfo{pages}{022501} (\bibinfo{year}{2012}).

\bibitem[{\citenamefont{Filter et~al.}(2012)\citenamefont{Filter, M\"uhlig,
  Eichelkraut, Rockstuhl, and Lederer}}]{Filter2012}
\bibinfo{author}{\bibfnamefont{R.}~\bibnamefont{Filter}},
  \bibinfo{author}{\bibfnamefont{S.}~\bibnamefont{M\"uhlig}},
  \bibinfo{author}{\bibfnamefont{T.}~\bibnamefont{Eichelkraut}},
  \bibinfo{author}{\bibfnamefont{C.}~\bibnamefont{Rockstuhl}},
  \bibnamefont{and} \bibinfo{author}{\bibfnamefont{F.}~\bibnamefont{Lederer}},
  \bibinfo{journal}{Phys. Rev. B} \textbf{\bibinfo{volume}{86}},
  \bibinfo{pages}{035404} (\bibinfo{year}{2012}).

\bibitem[{\citenamefont{Yoshie et~al.}(2004)\citenamefont{Yoshie, Scherer,
  Hendrickson, Khitrova, Gibbs, Rupper, Ell, Shchekin, and Deppe}}]{Yoshie2004}
\bibinfo{author}{\bibfnamefont{T.}~\bibnamefont{Yoshie}},
  \bibinfo{author}{\bibfnamefont{A.}~\bibnamefont{Scherer}},
  \bibinfo{author}{\bibfnamefont{J.}~\bibnamefont{Hendrickson}},
  \bibinfo{author}{\bibfnamefont{G.}~\bibnamefont{Khitrova}},
  \bibinfo{author}{\bibfnamefont{H.~M.} \bibnamefont{Gibbs}},
  \bibinfo{author}{\bibfnamefont{G.}~\bibnamefont{Rupper}},
  \bibinfo{author}{\bibfnamefont{C.}~\bibnamefont{Ell}},
  \bibinfo{author}{\bibfnamefont{O.~B.} \bibnamefont{Shchekin}},
  \bibnamefont{and} \bibinfo{author}{\bibfnamefont{D.~G.} \bibnamefont{Deppe}},
  \bibinfo{journal}{Nature} \textbf{\bibinfo{volume}{432}},
  \bibinfo{pages}{200} (\bibinfo{year}{2004}).

\bibitem[{\citenamefont{Reithmaier et~al.}(2004)\citenamefont{Reithmaier,
  S\k{e}k, L{\"o}ffler, Hofmann, Kuhn, Reitzenstein, Keldysh, Kulakovskii,
  Reinecke, and Forchel}}]{Reithmaier2004}
\bibinfo{author}{\bibfnamefont{J.~P.} \bibnamefont{Reithmaier}},
  \bibinfo{author}{\bibfnamefont{G.}~\bibnamefont{S\k{e}k}},
  \bibinfo{author}{\bibfnamefont{A.}~\bibnamefont{L{\"o}ffler}},
  \bibinfo{author}{\bibfnamefont{C.}~\bibnamefont{Hofmann}},
  \bibinfo{author}{\bibfnamefont{S.}~\bibnamefont{Kuhn}},
  \bibinfo{author}{\bibfnamefont{S.}~\bibnamefont{Reitzenstein}},
  \bibinfo{author}{\bibfnamefont{L.~V.} \bibnamefont{Keldysh}},
  \bibinfo{author}{\bibfnamefont{V.~D.} \bibnamefont{Kulakovskii}},
  \bibinfo{author}{\bibfnamefont{T.~L.} \bibnamefont{Reinecke}},
  \bibnamefont{and} \bibinfo{author}{\bibfnamefont{A.}~\bibnamefont{Forchel}},
  \bibinfo{journal}{Nature} \textbf{\bibinfo{volume}{432}},
  \bibinfo{pages}{197} (\bibinfo{year}{2004}).

\bibitem[{\citenamefont{Aoki et~al.}(2006)\citenamefont{Aoki, Dayan, Wilcut,
  Bowen, Parkins, Kippenberg, Vahala, and Kimble}}]{Aoki2006}
\bibinfo{author}{\bibfnamefont{T.}~\bibnamefont{Aoki}},
  \bibinfo{author}{\bibfnamefont{B.}~\bibnamefont{Dayan}},
  \bibinfo{author}{\bibfnamefont{E.}~\bibnamefont{Wilcut}},
  \bibinfo{author}{\bibfnamefont{W.}~\bibnamefont{Bowen}},
  \bibinfo{author}{\bibfnamefont{A.}~\bibnamefont{Parkins}},
  \bibinfo{author}{\bibfnamefont{T.~J.} \bibnamefont{Kippenberg}},
  \bibinfo{author}{\bibfnamefont{K.~J.} \bibnamefont{Vahala}},
  \bibnamefont{and} \bibinfo{author}{\bibfnamefont{H.~J.}
  \bibnamefont{Kimble}}, \bibinfo{journal}{Nature}
  \textbf{\bibinfo{volume}{443}}, \bibinfo{pages}{671} (\bibinfo{year}{2006}).

\bibitem[{\citenamefont{Yan et~al.}(2008)\citenamefont{Yan, Zhang, Duan, Zhao,
  and Govorov}}]{Yan2008}
\bibinfo{author}{\bibfnamefont{J.~Y.} \bibnamefont{Yan}},
  \bibinfo{author}{\bibfnamefont{W.}~\bibnamefont{Zhang}},
  \bibinfo{author}{\bibfnamefont{S.}~\bibnamefont{Duan}},
  \bibinfo{author}{\bibfnamefont{X.~G.} \bibnamefont{Zhao}}, \bibnamefont{and}
  \bibinfo{author}{\bibfnamefont{A.~O.} \bibnamefont{Govorov}},
  \bibinfo{journal}{Phys. Rev. B} \textbf{\bibinfo{volume}{77}},
  \bibinfo{pages}{165301} (\bibinfo{year}{2008}).

\bibitem[{\citenamefont{Artuso et~al.}(2011)\citenamefont{Artuso, Bryant,
  Garcia-Etxarri, and Aizpurua}}]{Artuso2011}
\bibinfo{author}{\bibfnamefont{R.~D.} \bibnamefont{Artuso}},
  \bibinfo{author}{\bibfnamefont{G.~W.} \bibnamefont{Bryant}},
  \bibinfo{author}{\bibfnamefont{A.}~\bibnamefont{Garcia-Etxarri}},
  \bibnamefont{and} \bibinfo{author}{\bibfnamefont{J.}~\bibnamefont{Aizpurua}},
  \bibinfo{journal}{Phys. Rev. B} \textbf{\bibinfo{volume}{83}},
  \bibinfo{pages}{235406} (\bibinfo{year}{2011}).

\bibitem[{\citenamefont{Meystre and Sargent}(1999)}]{Meystre1999}
\bibinfo{author}{\bibfnamefont{P.}~\bibnamefont{Meystre}} \bibnamefont{and}
  \bibinfo{author}{\bibfnamefont{M.}~\bibnamefont{Sargent}},
  \emph{\bibinfo{title}{Elements of Quantum Optics}}
  (\bibinfo{publisher}{Springer Verlag}, \bibinfo{year}{1999}).

\bibitem[{\citenamefont{Zuloaga et~al.}(2010)\citenamefont{Zuloaga, Prodan, and
  Nordlander}}]{Zuloaga2010}
\bibinfo{author}{\bibfnamefont{J.}~\bibnamefont{Zuloaga}},
  \bibinfo{author}{\bibfnamefont{E.}~\bibnamefont{Prodan}}, \bibnamefont{and}
  \bibinfo{author}{\bibfnamefont{P.}~\bibnamefont{Nordlander}},
  \bibinfo{journal}{ACS Nano} \textbf{\bibinfo{volume}{4}},
  \bibinfo{pages}{5269} (\bibinfo{year}{2010}).

\bibitem[{\citenamefont{Manjavacas et~al.}(2011)\citenamefont{Manjavacas,
  Garcia~de Abajo, and Nordlander}}]{Manjavacas2011}
\bibinfo{author}{\bibfnamefont{A.}~\bibnamefont{Manjavacas}},
  \bibinfo{author}{\bibfnamefont{F.~J.} \bibnamefont{Garcia~de Abajo}},
  \bibnamefont{and}
  \bibinfo{author}{\bibfnamefont{P.}~\bibnamefont{Nordlander}},
  \bibinfo{journal}{Nano Lett.} \textbf{\bibinfo{volume}{11}},
  \bibinfo{pages}{2318} (\bibinfo{year}{2011}).

\bibitem[{\citenamefont{Tr\"ugler and Hohenester}(2008)}]{Hohenester2008}
\bibinfo{author}{\bibfnamefont{A.}~\bibnamefont{Tr\"ugler}} \bibnamefont{and}
  \bibinfo{author}{\bibfnamefont{U.}~\bibnamefont{Hohenester}},
  \bibinfo{journal}{Phys. Rev. B} \textbf{\bibinfo{volume}{77}},
  \bibinfo{pages}{115403} (\bibinfo{year}{2008}).

\bibitem[{\citenamefont{Waks and Sridharan}(2010)}]{Waks2010}
\bibinfo{author}{\bibfnamefont{E.}~\bibnamefont{Waks}} \bibnamefont{and}
  \bibinfo{author}{\bibfnamefont{D.}~\bibnamefont{Sridharan}},
  \bibinfo{journal}{Phys. Rev. A} \textbf{\bibinfo{volume}{82}},
  \bibinfo{pages}{043845} (\bibinfo{year}{2010}).

\bibitem[{\citenamefont{Dzsotjan et~al.}(2011)\citenamefont{Dzsotjan,
  K{\"a}stel, and Fleischhauer}}]{Dzsotjan2011}
\bibinfo{author}{\bibfnamefont{D.}~\bibnamefont{Dzsotjan}},
  \bibinfo{author}{\bibfnamefont{J.}~\bibnamefont{K{\"a}stel}},
  \bibnamefont{and}
  \bibinfo{author}{\bibfnamefont{M.}~\bibnamefont{Fleischhauer}},
  \bibinfo{journal}{Phys. Rev. B} \textbf{\bibinfo{volume}{84}},
  \bibinfo{pages}{075419} (\bibinfo{year}{2011}).

\bibitem[{\citenamefont{Zhang and Govorov}(2011)}]{GovorovQuantumFano}
\bibinfo{author}{\bibfnamefont{W.}~\bibnamefont{Zhang}} \bibnamefont{and}
  \bibinfo{author}{\bibfnamefont{A.~O.} \bibnamefont{Govorov}},
  \bibinfo{journal}{Phys. Rev. B} \textbf{\bibinfo{volume}{84}},
  \bibinfo{pages}{081405} (\bibinfo{year}{2011}).

\bibitem[{\citenamefont{H\"ummer et~al.}(2013)\citenamefont{H\"ummer,
  Garcia-Vidal, Martin-Moreno, and Zueco}}]{Hummer2013}
\bibinfo{author}{\bibfnamefont{T.}~\bibnamefont{H\"ummer}},
  \bibinfo{author}{\bibfnamefont{F.~J.} \bibnamefont{Garcia-Vidal}},
  \bibinfo{author}{\bibfnamefont{L.}~\bibnamefont{Martin-Moreno}},
  \bibnamefont{and} \bibinfo{author}{\bibfnamefont{D.}~\bibnamefont{Zueco}},
  \bibinfo{journal}{Phys. Rev. B} \textbf{\bibinfo{volume}{87}},
  \bibinfo{pages}{115419} (\bibinfo{year}{2013}).

\bibitem[{\citenamefont{Wolters et~al.}(2012)\citenamefont{Wolters, Kewes,
  Schell, N{\"u}sse, Schoengen, L{\"o}chel, Hanke, Bratschitsch, Leitenstorfer,
  Aichele et~al.}}]{Benson2012}
\bibinfo{author}{\bibfnamefont{J.}~\bibnamefont{Wolters}},
  \bibinfo{author}{\bibfnamefont{G.}~\bibnamefont{Kewes}},
  \bibinfo{author}{\bibfnamefont{A.~W.} \bibnamefont{Schell}},
  \bibinfo{author}{\bibfnamefont{N.}~\bibnamefont{N{\"u}sse}},
  \bibinfo{author}{\bibfnamefont{M.}~\bibnamefont{Schoengen}},
  \bibinfo{author}{\bibfnamefont{B.}~\bibnamefont{L{\"o}chel}},
  \bibinfo{author}{\bibfnamefont{T.}~\bibnamefont{Hanke}},
  \bibinfo{author}{\bibfnamefont{R.}~\bibnamefont{Bratschitsch}},
  \bibinfo{author}{\bibfnamefont{A.}~\bibnamefont{Leitenstorfer}},
  \bibinfo{author}{\bibfnamefont{T.}~\bibnamefont{Aichele}},
  \bibnamefont{et~al.}, \bibinfo{journal}{Phys. Status Solidi B}
  \textbf{\bibinfo{volume}{249}}, \bibinfo{pages}{918} (\bibinfo{year}{2012}).

\bibitem[{\citenamefont{S{\l}owik et~al.}(2012)\citenamefont{S{\l}owik,
  Raczy\'{n}ski, Zaremba, and Zieli\'{n}ska-Kaniasty}}]{Karolina2012}
\bibinfo{author}{\bibfnamefont{K.}~\bibnamefont{S{\l}owik}},
  \bibinfo{author}{\bibfnamefont{A.}~\bibnamefont{Raczy\'{n}ski}},
  \bibinfo{author}{\bibfnamefont{J.}~\bibnamefont{Zaremba}}, \bibnamefont{and}
  \bibinfo{author}{\bibfnamefont{S.}~\bibnamefont{Zieli\'{n}ska-Kaniasty}},
  \bibinfo{journal}{Optics Communications} \textbf{\bibinfo{volume}{285}},
  \bibinfo{pages}{2392} (\bibinfo{year}{2012}).

\bibitem[{\citenamefont{Lembessis and Babiker}(2013)}]{Lembessis2013}
\bibinfo{author}{\bibfnamefont{V.~E.} \bibnamefont{Lembessis}}
  \bibnamefont{and} \bibinfo{author}{\bibfnamefont{M.}~\bibnamefont{Babiker}},
  \bibinfo{journal}{Phys. Rev. Lett.} \textbf{\bibinfo{volume}{110}},
  \bibinfo{pages}{083002} (\bibinfo{year}{2013}).

\bibitem[{\citenamefont{Heiss et~al.}(2007)\citenamefont{Heiss, Schaeck, Huebl,
  Bichler, Abstreiter, Finley, Bulaev, and Loss}}]{Heiss2007}
\bibinfo{author}{\bibfnamefont{D.}~\bibnamefont{Heiss}},
  \bibinfo{author}{\bibfnamefont{S.}~\bibnamefont{Schaeck}},
  \bibinfo{author}{\bibfnamefont{H.}~\bibnamefont{Huebl}},
  \bibinfo{author}{\bibfnamefont{M.}~\bibnamefont{Bichler}},
  \bibinfo{author}{\bibfnamefont{G.}~\bibnamefont{Abstreiter}},
  \bibinfo{author}{\bibfnamefont{J.~J.} \bibnamefont{Finley}},
  \bibinfo{author}{\bibfnamefont{D.~V.} \bibnamefont{Bulaev}},
  \bibnamefont{and} \bibinfo{author}{\bibfnamefont{D.}~\bibnamefont{Loss}},
  \bibinfo{journal}{Phys. Rev. B} \textbf{\bibinfo{volume}{76}},
  \bibinfo{pages}{241306} (\bibinfo{year}{2007}).

\bibitem[{\citenamefont{Chen et~al.}(2013)\citenamefont{Chen, Sandoghdar, and
  Agio}}]{Chen2012coherent}
\bibinfo{author}{\bibfnamefont{X.-W.} \bibnamefont{Chen}},
  \bibinfo{author}{\bibfnamefont{V.}~\bibnamefont{Sandoghdar}},
  \bibnamefont{and} \bibinfo{author}{\bibfnamefont{M.}~\bibnamefont{Agio}},
  \bibinfo{journal}{Phys. Rev. Lett.} \textbf{\bibinfo{volume}{110}},
  \bibinfo{pages}{153605} (\bibinfo{year}{2013}).

\bibitem[{\citenamefont{Dorner and Zoller}(2002)}]{Dorner2002}
\bibinfo{author}{\bibfnamefont{U.}~\bibnamefont{Dorner}} \bibnamefont{and}
  \bibinfo{author}{\bibfnamefont{P.}~\bibnamefont{Zoller}},
  \bibinfo{journal}{Phys. Rev. A} \textbf{\bibinfo{volume}{66}},
  \bibinfo{pages}{023816} (\bibinfo{year}{2002}).

\bibitem[{\citenamefont{Wolfram}(1999)}]{wolfram1999mathematica}
\bibinfo{author}{\bibfnamefont{S.}~\bibnamefont{Wolfram}},
  \emph{\bibinfo{title}{The MATHEMATICA Book, Version 4}}
  (\bibinfo{publisher}{Cambr. Univ. Pr.}, \bibinfo{year}{1999}).

\bibitem[{\citenamefont{Wallraff et~al.}(2004)\citenamefont{Wallraff, Schuster,
  Blais, Frunzio, Huang, Majer, Kumar, Girvin, and
  Schoelkopf}}]{wallraff2004strong}
\bibinfo{author}{\bibfnamefont{A.}~\bibnamefont{Wallraff}},
  \bibinfo{author}{\bibfnamefont{D.~I.} \bibnamefont{Schuster}},
  \bibinfo{author}{\bibfnamefont{A.}~\bibnamefont{Blais}},
  \bibinfo{author}{\bibfnamefont{L.}~\bibnamefont{Frunzio}},
  \bibinfo{author}{\bibfnamefont{R.-S.} \bibnamefont{Huang}},
  \bibinfo{author}{\bibfnamefont{J.}~\bibnamefont{Majer}},
  \bibinfo{author}{\bibfnamefont{S.}~\bibnamefont{Kumar}},
  \bibinfo{author}{\bibfnamefont{S.~M.} \bibnamefont{Girvin}},
  \bibnamefont{and} \bibinfo{author}{\bibfnamefont{R.~J.}
  \bibnamefont{Schoelkopf}}, \bibinfo{journal}{Nature}
  \textbf{\bibinfo{volume}{431}}, \bibinfo{pages}{162} (\bibinfo{year}{2004}).

\bibitem[{\citenamefont{Andreani et~al.}(1999)\citenamefont{Andreani,
  Panzarini, and G\'erard}}]{Andreani1999}
\bibinfo{author}{\bibfnamefont{L.~C.} \bibnamefont{Andreani}},
  \bibinfo{author}{\bibfnamefont{G.}~\bibnamefont{Panzarini}},
  \bibnamefont{and} \bibinfo{author}{\bibfnamefont{J.-M.}
  \bibnamefont{G\'erard}}, \bibinfo{journal}{Phys. Rev. B}
  \textbf{\bibinfo{volume}{60}}, \bibinfo{pages}{13276} (\bibinfo{year}{1999}).

\bibitem[{\citenamefont{Savasta et~al.}(2010)\citenamefont{Savasta, Saija,
  Ridolfo, Di~Stefano, Denti, and Borghese}}]{savasta2010nanopolaritons}
\bibinfo{author}{\bibfnamefont{S.}~\bibnamefont{Savasta}},
  \bibinfo{author}{\bibfnamefont{R.}~\bibnamefont{Saija}},
  \bibinfo{author}{\bibfnamefont{A.}~\bibnamefont{Ridolfo}},
  \bibinfo{author}{\bibfnamefont{O.}~\bibnamefont{Di~Stefano}},
  \bibinfo{author}{\bibfnamefont{P.}~\bibnamefont{Denti}}, \bibnamefont{and}
  \bibinfo{author}{\bibfnamefont{F.}~\bibnamefont{Borghese}},
  \bibinfo{journal}{ACS Nano} \textbf{\bibinfo{volume}{4}},
  \bibinfo{pages}{6369} (\bibinfo{year}{2010}).

\bibitem[{\citenamefont{Novotny and Hecht}(2006)}]{NovotnyHecht}
\bibinfo{author}{\bibfnamefont{L.}~\bibnamefont{Novotny}} \bibnamefont{and}
  \bibinfo{author}{\bibfnamefont{B.}~\bibnamefont{Hecht}},
  \emph{\bibinfo{title}{Principles Of Nano-Optics}}
  (\bibinfo{publisher}{Cambridge University Press}, \bibinfo{year}{2006}).

\bibitem[{\citenamefont{Bharadwaj et~al.}(2009)\citenamefont{Bharadwaj,
  Deutsch, and Novotny}}]{Bharadwaj2009}
\bibinfo{author}{\bibfnamefont{P.}~\bibnamefont{Bharadwaj}},
  \bibinfo{author}{\bibfnamefont{B.}~\bibnamefont{Deutsch}}, \bibnamefont{and}
  \bibinfo{author}{\bibfnamefont{L.}~\bibnamefont{Novotny}},
  \bibinfo{journal}{Adv. Opt. Photon.} \textbf{\bibinfo{volume}{1}},
  \bibinfo{pages}{438} (\bibinfo{year}{2009}).

\bibitem[{\citenamefont{Palik}(1985)}]{palik}
\bibinfo{author}{\bibfnamefont{E.~D.} \bibnamefont{Palik}},
  \emph{\bibinfo{title}{Handbook of Optical Constants of Solids}}, no.
  \bibinfo{number}{Bd. 1} in \bibinfo{series}{Handbook of Optical Constants of
  Solids, Five-Volume Set} (\bibinfo{publisher}{Elsevier Science},
  \bibinfo{year}{1985}).

\bibitem[{\citenamefont{Ge et~al.}(2013)\citenamefont{Ge, Van~Vlack, Yao,
  Young, and Hughes}}]{Ge2013}
\bibinfo{author}{\bibfnamefont{R.-C.} \bibnamefont{Ge}},
  \bibinfo{author}{\bibfnamefont{C.}~\bibnamefont{Van~Vlack}},
  \bibinfo{author}{\bibfnamefont{P.}~\bibnamefont{Yao}},
  \bibinfo{author}{\bibfnamefont{J.~F.} \bibnamefont{Young}}, \bibnamefont{and}
  \bibinfo{author}{\bibfnamefont{S.}~\bibnamefont{Hughes}},
  \bibinfo{journal}{Phys. Rev. B} \textbf{\bibinfo{volume}{87}},
  \bibinfo{pages}{205425} (\bibinfo{year}{2013}).

\bibitem[{\citenamefont{Mu and Savage}(1992)}]{Mu1992}
\bibinfo{author}{\bibfnamefont{Y.}~\bibnamefont{Mu}} \bibnamefont{and}
  \bibinfo{author}{\bibfnamefont{C.~M.} \bibnamefont{Savage}},
  \bibinfo{journal}{Phys. Rev. A} \textbf{\bibinfo{volume}{46}},
  \bibinfo{pages}{5944} (\bibinfo{year}{1992}).

\bibitem[{\citenamefont{Maksymov et~al.}(2012)\citenamefont{Maksymov,
  Miroshnichenko, and Kivshar}}]{Maksymov2012}
\bibinfo{author}{\bibfnamefont{I.~S.} \bibnamefont{Maksymov}},
  \bibinfo{author}{\bibfnamefont{A.~E.} \bibnamefont{Miroshnichenko}},
  \bibnamefont{and} \bibinfo{author}{\bibfnamefont{Y.~S.}
  \bibnamefont{Kivshar}}, \bibinfo{journal}{Phys. Rev. A}
  \textbf{\bibinfo{volume}{86}}, \bibinfo{pages}{011801}
  (\bibinfo{year}{2012}).

\bibitem[{\citenamefont{Bohren and Huffman}(1998)}]{Bohren1998}
\bibinfo{author}{\bibfnamefont{C.}~\bibnamefont{Bohren}} \bibnamefont{and}
  \bibinfo{author}{\bibfnamefont{D.~R.} \bibnamefont{Huffman}},
  \emph{\bibinfo{title}{Absorption and Scattering of Light by Small Particles}}
  (\bibinfo{publisher}{Wiley Science Paperback Series}, \bibinfo{year}{1998}).

\bibitem[{\citenamefont{Gonzalez-Tudela
  et~al.}(2011)\citenamefont{Gonzalez-Tudela, Martin-Cano, Moreno,
  Martin-Moreno, Tejedor, and Garcia-Vidal}}]{Gonzalez}
\bibinfo{author}{\bibfnamefont{A.}~\bibnamefont{Gonzalez-Tudela}},
  \bibinfo{author}{\bibfnamefont{D.}~\bibnamefont{Martin-Cano}},
  \bibinfo{author}{\bibfnamefont{E.}~\bibnamefont{Moreno}},
  \bibinfo{author}{\bibfnamefont{L.}~\bibnamefont{Martin-Moreno}},
  \bibinfo{author}{\bibfnamefont{C.}~\bibnamefont{Tejedor}}, \bibnamefont{and}
  \bibinfo{author}{\bibfnamefont{F.~J.} \bibnamefont{Garcia-Vidal}},
  \bibinfo{journal}{Phys. Rev. Lett.} \textbf{\bibinfo{volume}{106}},
  \bibinfo{pages}{020501} (\bibinfo{year}{2011}).

\bibitem[{\citenamefont{Martin-Cano et~al.}(2011)\citenamefont{Martin-Cano,
  Gonz\'{a}lez-Tudela, Martin-Moreno, Garcia-Vidal, Tejedor, and
  Moreno}}]{Diego}
\bibinfo{author}{\bibfnamefont{D.}~\bibnamefont{Martin-Cano}},
  \bibinfo{author}{\bibfnamefont{A.}~\bibnamefont{Gonz\'{a}lez-Tudela}},
  \bibinfo{author}{\bibfnamefont{L.}~\bibnamefont{Martin-Moreno}},
  \bibinfo{author}{\bibfnamefont{F.~J.} \bibnamefont{Garcia-Vidal}},
  \bibinfo{author}{\bibfnamefont{C.}~\bibnamefont{Tejedor}}, \bibnamefont{and}
  \bibinfo{author}{\bibfnamefont{E.}~\bibnamefont{Moreno}},
  \bibinfo{journal}{Phys. Rev. B} \textbf{\bibinfo{volume}{84}},
  \bibinfo{pages}{235306} (\bibinfo{year}{2011}).

\bibitem[{\citenamefont{Sidorkin et~al.}(2009)\citenamefont{Sidorkin, van
  Veldhoven, van~der Drift, Alkemade, Salemink, and Maas}}]{sidorkin2009sub}
\bibinfo{author}{\bibfnamefont{V.}~\bibnamefont{Sidorkin}},
  \bibinfo{author}{\bibfnamefont{E.}~\bibnamefont{van Veldhoven}},
  \bibinfo{author}{\bibfnamefont{E.}~\bibnamefont{van~der Drift}},
  \bibinfo{author}{\bibfnamefont{P.}~\bibnamefont{Alkemade}},
  \bibinfo{author}{\bibfnamefont{H.}~\bibnamefont{Salemink}}, \bibnamefont{and}
  \bibinfo{author}{\bibfnamefont{D.}~\bibnamefont{Maas}}, \bibinfo{journal}{J.
  Vac. Sci. Technol. B: Microelectronics and Nanometer Structures}
  \textbf{\bibinfo{volume}{27}}, \bibinfo{pages}{L18} (\bibinfo{year}{2009}).

\bibitem[{\citenamefont{Bleuse et~al.}(2011)\citenamefont{Bleuse, Claudon,
  Creasey, Malik, G\'erard, Maksymov, Hugonin, and Lalanne}}]{Bleuse2011}
\bibinfo{author}{\bibfnamefont{J.}~\bibnamefont{Bleuse}},
  \bibinfo{author}{\bibfnamefont{J.}~\bibnamefont{Claudon}},
  \bibinfo{author}{\bibfnamefont{M.}~\bibnamefont{Creasey}},
  \bibinfo{author}{\bibfnamefont{N.~S.} \bibnamefont{Malik}},
  \bibinfo{author}{\bibfnamefont{J.-M.} \bibnamefont{G\'erard}},
  \bibinfo{author}{\bibfnamefont{I.}~\bibnamefont{Maksymov}},
  \bibinfo{author}{\bibfnamefont{J.-P.} \bibnamefont{Hugonin}},
  \bibnamefont{and} \bibinfo{author}{\bibfnamefont{P.}~\bibnamefont{Lalanne}},
  \bibinfo{journal}{Phys. Rev. Lett.} \textbf{\bibinfo{volume}{106}},
  \bibinfo{pages}{103601} (\bibinfo{year}{2011}).

\bibitem[{\citenamefont{Chen et~al.}(2012)\citenamefont{Chen, Chen, Lai, and
  Li}}]{chen2012single}
\bibinfo{author}{\bibfnamefont{I.-H.} \bibnamefont{Chen}},
  \bibinfo{author}{\bibfnamefont{K.-H.} \bibnamefont{Chen}},
  \bibinfo{author}{\bibfnamefont{W.-T.} \bibnamefont{Lai}}, \bibnamefont{and}
  \bibinfo{author}{\bibfnamefont{P.-W.} \bibnamefont{Li}},
  \bibinfo{journal}{IEEE Transactions on Electron Devices}
  \textbf{\bibinfo{volume}{59}}, \bibinfo{pages}{3224} (\bibinfo{year}{2012}).

\bibitem[{\citenamefont{Alaee et~al.}(2013)\citenamefont{Alaee, Menzel,
  Huebner, Pshenay-Severin, Bin~Hasan, Pertsch, Rockstuhl, and
  Lederer}}]{Alaee2013deep}
\bibinfo{author}{\bibfnamefont{R.}~\bibnamefont{Alaee}},
  \bibinfo{author}{\bibfnamefont{C.}~\bibnamefont{Menzel}},
  \bibinfo{author}{\bibfnamefont{U.}~\bibnamefont{Huebner}},
  \bibinfo{author}{\bibfnamefont{E.}~\bibnamefont{Pshenay-Severin}},
  \bibinfo{author}{\bibfnamefont{S.}~\bibnamefont{Bin~Hasan}},
  \bibinfo{author}{\bibfnamefont{T.}~\bibnamefont{Pertsch}},
  \bibinfo{author}{\bibfnamefont{C.}~\bibnamefont{Rockstuhl}},
  \bibnamefont{and} \bibinfo{author}{\bibfnamefont{F.}~\bibnamefont{Lederer}},
  \bibinfo{journal}{Nano Lett.} \textbf{\bibinfo{volume}{13}},
  \bibinfo{pages}{3482} (\bibinfo{year}{2013}).

\bibitem[{\citenamefont{Chang et~al.}(2006)\citenamefont{Chang, S\o{}rensen,
  Hemmer, and Lukin}}]{TheLukinPaper}
\bibinfo{author}{\bibfnamefont{D.~E.} \bibnamefont{Chang}},
  \bibinfo{author}{\bibfnamefont{A.~S.} \bibnamefont{S\o{}rensen}},
  \bibinfo{author}{\bibfnamefont{P.~R.} \bibnamefont{Hemmer}},
  \bibnamefont{and} \bibinfo{author}{\bibfnamefont{M.~D.} \bibnamefont{Lukin}},
  \bibinfo{journal}{Phys. Rev. Lett.} \textbf{\bibinfo{volume}{97}},
  \bibinfo{pages}{053002} (\bibinfo{year}{2006}).

\bibitem[{\citenamefont{Tan}(1999)}]{tan1999computational}
\bibinfo{author}{\bibfnamefont{S.~M.} \bibnamefont{Tan}}, \bibinfo{journal}{J.
  Opt. B} \textbf{\bibinfo{volume}{1}}, \bibinfo{pages}{424}
  (\bibinfo{year}{1999}).

\bibitem[{\citenamefont{Hood et~al.}(1998)\citenamefont{Hood, Chapman, Lynn,
  and Kimble}}]{Hood1998}
\bibinfo{author}{\bibfnamefont{C.~J.} \bibnamefont{Hood}},
  \bibinfo{author}{\bibfnamefont{M.~S.} \bibnamefont{Chapman}},
  \bibinfo{author}{\bibfnamefont{T.~W.} \bibnamefont{Lynn}}, \bibnamefont{and}
  \bibinfo{author}{\bibfnamefont{H.~J.} \bibnamefont{Kimble}},
  \bibinfo{journal}{Phys. Rev. Lett.} \textbf{\bibinfo{volume}{80}},
  \bibinfo{pages}{4157} (\bibinfo{year}{1998}).

\bibitem[{\citenamefont{Miller et~al.}(2005)\citenamefont{Miller, Northup,
  Birnbaum, Boca, Boozer, and Kimble}}]{Miller2005}
\bibinfo{author}{\bibfnamefont{R.}~\bibnamefont{Miller}},
  \bibinfo{author}{\bibfnamefont{T.~E.} \bibnamefont{Northup}},
  \bibinfo{author}{\bibfnamefont{K.~M.} \bibnamefont{Birnbaum}},
  \bibinfo{author}{\bibfnamefont{A.}~\bibnamefont{Boca}},
  \bibinfo{author}{\bibfnamefont{A.~D.} \bibnamefont{Boozer}},
  \bibnamefont{and} \bibinfo{author}{\bibfnamefont{H.~J.}
  \bibnamefont{Kimble}}, \bibinfo{journal}{J. Phys. B: At. Mol. Opt. Phys.}
  \textbf{\bibinfo{volume}{38}}, \bibinfo{pages}{S551} (\bibinfo{year}{2005}).

\bibitem[{\citenamefont{Petschulat et~al.}(2008)\citenamefont{Petschulat,
  Menzel, Chipouline, Rockstuhl, Tuennermann, Lederer, and
  Pertsch}}]{Petschulat2008}
\bibinfo{author}{\bibfnamefont{J.}~\bibnamefont{Petschulat}},
  \bibinfo{author}{\bibfnamefont{C.}~\bibnamefont{Menzel}},
  \bibinfo{author}{\bibfnamefont{A.}~\bibnamefont{Chipouline}},
  \bibinfo{author}{\bibfnamefont{C.}~\bibnamefont{Rockstuhl}},
  \bibinfo{author}{\bibfnamefont{A.}~\bibnamefont{Tuennermann}},
  \bibinfo{author}{\bibfnamefont{F.}~\bibnamefont{Lederer}}, \bibnamefont{and}
  \bibinfo{author}{\bibfnamefont{T.}~\bibnamefont{Pertsch}},
  \bibinfo{journal}{Phys. Rev. A} \textbf{\bibinfo{volume}{78}},
  \bibinfo{pages}{043811} (\bibinfo{year}{2008}).

\bibitem[{\citenamefont{Landau et~al.}(1982)\citenamefont{Landau, Lifshitz, and
  Pitaevskii}}]{Landau}
\bibinfo{author}{\bibfnamefont{L.~D.} \bibnamefont{Landau}},
  \bibinfo{author}{\bibfnamefont{E.~M.} \bibnamefont{Lifshitz}},
  \bibnamefont{and} \bibinfo{author}{\bibfnamefont{L.~P.}
  \bibnamefont{Pitaevskii}}, \emph{\bibinfo{title}{Electrodynamics of Continous
  Media; Landau and Lifshitz Course of Theoretical Physics}}, vol.
  \bibinfo{volume}{2nd} (\bibinfo{publisher}{Butterworth-Heinenann},
  \bibinfo{address}{Boston, MA}, \bibinfo{year}{1982}).

\bibitem[{\citenamefont{Dicke}(1954)}]{Dicke1954}
\bibinfo{author}{\bibfnamefont{R.~H.} \bibnamefont{Dicke}},
  \bibinfo{journal}{Phys. Rev.} \textbf{\bibinfo{volume}{93}},
  \bibinfo{pages}{99} (\bibinfo{year}{1954}).

\end{thebibliography}

\end{document}